\documentclass[nofootinbib,aps,amssymb,floatfix,preprintnumbers,groupedaddress]{revtex4}

\usepackage[T1]{fontenc}
\usepackage{blindtext}

\usepackage{graphics,graphicx}
\usepackage{dcolumn}
\usepackage{bm}
\usepackage{epsfig}
\usepackage[usenames]{color}
\usepackage[hidelinks]{hyperref} 
\usepackage{epstopdf}
\usepackage{simplewick}
\usepackage[utf8]{inputenc} 
\usepackage{slashed}
\usepackage{soul}
\usepackage[dvipsnames]{xcolor}
\usepackage[export]{adjustbox}
\usepackage{amsmath}
\usepackage{amssymb} 
\usepackage{bm}
\usepackage{latexsym}
\usepackage{amsfonts} 

\begin{document}

\title{Numerical study of the twist-3 asymmetry $\boldsymbol{A_{LT}}$ in single-inclusive \\ electron-nucleon and proton-proton collisions}

\author{Brandon Bauer} 
\author{Daniel Pitonyak} 
\author{Cody Shay\vspace{0.15cm}} 

\affiliation{Department of Physics, Lebanon Valley College, Annville, PA 17003, USA}

\begin{abstract}
We provide the first rigorous numerical analysis of the longitudinal-transverse double-spin asymmetry $A_{LT}$ in electron-nucleon and proton-proton collisions for the case where only a single pion, jet, or photon is detected in the final state.  Given recent extractions of certain, previously unknown, non-perturbative functions, we are able to compute contributions from all terms relevant for $A_{LT}$ and make realistic predictions for the observable at Jefferson Lab (JLab) 12 GeV, COMPASS, the future Electron-Ion Collider, and the Relativistic Heavy Ion Collider.  We also compare our results to a JLab 6 GeV measurement, which are the only data available for this type of reaction.  The twist-3 nature of $A_{LT}$  makes it a potentially fruitful avenue to probe quark-gluon-quark correlations in hadrons as well as  provide insights into dynamical quark mass generation in QCD.
\end{abstract}

\maketitle

\section{Introduction}
\label{sec:intro}
One of the earliest puzzles in spin physics research was the observation in the 1970s of large asymmetries in single-inclusive reactions where one hadron is transversely polarized~\cite{Bunce:1976yb,Klem:1976ui} -- so-called single transverse-spin asymmetries (SSAs) $A_N$.  This eventually was recognized as a signature of multi-parton correlations in hadrons~\cite{Efremov:1981sh,Efremov:1984ip,Qiu:1991pp,Qiu:1991wg,Qiu:1998ia} and has been a source of intense theoretical~\cite{Efremov:1981sh,Efremov:1984ip,Qiu:1991pp,Qiu:1991wg,Qiu:1998ia,Kanazawa:2000hz,Eguchi:2006qz,Kouvaris:2006zy,Eguchi:2006mc,Zhou:2008fb,Koike:2009ge,Metz:2012ct,Kanazawa:2013uia,Beppu:2013uda,Kanazawa:2015ajw,Koike:2017fxr,Koike:2019zxc,Koike:2021awj,Koike:2022ddx,Ikarashi:2022yzg,Ikarashi:2022zeo}, phenomenological~\cite{Qiu:1998ia,Kanazawa:2000kp,Kouvaris:2006zy,Kanazawa:2010au,Kang:2011hk,Metz:2012ui,Beppu:2013uda,Gamberg:2013kla,Kanazawa:2014dca,Gamberg:2014eia,Gamberg:2017gle,Cammarota:2020qcw,Gamberg:2022kdb}, and experimental~\cite{Adams:1991rw,Krueger:1998hz,Allgower:2002qi,Adams:2003fx,Adler:2005in,Lee:2007zzh,Abelev:2008af,Arsene:2008aa,Adamczyk:2012qj,Adamczyk:2012xd,Bland:2013pkt,Adare:2013ekj,Adare:2014qzo,Airapetian:2013bim,Allada:2013nsw,STAR:2020nnl} activity for decades.  The collinear twist-3 formalism that underpins this work allows one to explore a rich set of non-perturbative functions, of which SSAs are sensitive to a certain subset.  Namely, the na\"{i}ve time-reversal odd (T-odd) nature of SSAs gives access to pole contributions from initial state multi-parton distribution functions (PDFs) (where typically one of the partons' momentum fractions vanishes~\cite{Qiu:1991pp,Qiu:1991wg,Qiu:1998ia,Kanazawa:2000hz,Kouvaris:2006zy,Koike:2009ge,Beppu:2013uda}\footnote{The poles are due to propagators in the hard scattering going on shell.  While usually this causes a momentum fraction in the multi-parton PDF to vanish (``soft poles''), there are certain processes that also lead to ``hard poles''~\cite{Eguchi:2006qz,Eguchi:2006mc,Albaltan:2019cyc}, where all parton momentum fractions remain nonzero.}); or to the imaginary part of (non-pole) final-state multi-parton fragmentation functions (FFs)~\cite{Metz:2012ct,Kanazawa:2013uia}.\footnote{We will still refer to initial-state twist-3 functions as parton distribution functions (PDFs) and final-state twist-3 functions as fragmentation functions (FFs), even though they do not have a strict probability interpretation.}  For example, $A_N$ in $p^\uparrow p\to \pi\,X$ at forward rapidity is mainly sensitive to the Qiu-Sterman PDF $F_{FT}(x,x)$ (where the two quarks carry the same momentum fraction $x$), as well as $H_1^{\perp(1)}(z)$ (which is the first-moment of the Collins function) and $\tilde{H}(z)$, with $z$ the momentum fraction carried by the produced hadron.  The latter two functions are certain integrals over $z_1$ (from $z$ to $\infty$) of the FF $\hat{H}^\Im_{FU}(z,z_1)$~\cite{Kanazawa:2015ajw}, where $\Im$ indicates the imaginary part. There are a plethora of SSA measurements, not only in $p^\uparrow p\to h\,X$ but also semi-inclusive deep-inelastic scattering (SIDIS) $e\,N^\uparrow\to e\,h\,X$~\cite{Airapetian:2009ae, Alekseev:2008aa,Airapetian:2010ds,Qian:2011py,Adolph:2014zba,Zhao:2014qvx,Adolph:2016dvl,HERMES:2020ifk}, electron-positron annihilation $e^+e^-\to h_1\,h_2\,X$~\cite{Seidl:2008xc,TheBABAR:2013yha,Aubert:2015hha,Ablikim:2015pta,Li:2019iyt}, and Drell-Yan $p^\uparrow p\to \{W^\pm,Z,\,{\rm or}\;\ell^+\ell^-\}\,X$~\cite{Adamczyk:2015gyk,Aghasyan:2017jop}.  Due to this data, as well as the connection between collinear twist-3 and transverse momentum dependent (TMD) functions~\cite{Ji:2006ub,Ji:2006br,Koike:2007dg,Yuan:2009dw,Zhou:2009jm}, $F_{FT}(x,x)$, $H_1^{\perp(1)}(z)$, and $\tilde{H}(z)$, along with the twist-2 transversity PDF $h_1(x)$, have all been extracted in various phenomenological analyses~(see, e.g., \cite{Kanazawa:2014dca,Echevarria:2014xaa,Kang:2015msa,Echevarria:2020hpy,Bury:2021sue,Cammarota:2020qcw,Gamberg:2022kdb}).

A complimentary observable to study multi-parton correlations in hadrons is the longitudinal-transverse double-spin asymmetry $A_{LT}$ in collisions like $\vec{e}\,N^\uparrow \to \pi\,X$ and $p^\uparrow \vec{p}\to \pi\,X$.  These are {\it T-even} reactions that are sensitive to the {\it non-pole} pieces of certain multi-parton PDFs (e.g., $F_{FT}(x,x_1)$ with $x\neq x_1$) and the {\it real part} $\Re$ of certain multi-parton FFs (e.g., $\hat{H}^\Re_{FU}(z,z_1)$). From the theoretical side, $A_{LT}$ has been well studied in electron-nucleon~\cite{Kang:2011jw,Kanazawa:2014tda,Kanazawa:2015ajw} and proton-proton~\cite{Liang:2012rb,Metz:2012fq,Hatta:2013wsa,Koike:2015yza,Koike:2016ura} collisions for various single-inclusive final states (e.g., hadron, jet, or photon), with some limited numerical work performed for the electron-nucleon case~\cite{Kang:2011jw,Kanazawa:2014tda}, but none for proton-proton.  The main hindrance to more rigorous predictions has been the lack of input for important non-perturbative functions in $A_{LT}$, which forces one to resort to approximations or the outright neglect of certain terms~\cite{Kang:2011jw,Kanazawa:2014tda}.  For example, one of the main PDFs that enters $A_{LT}$ is $g_{1T}^{(1)}(x)$, which is the first-moment of the worm-gear TMD $g_{1T}$, and it has only been extracted recently~\cite{Bhattacharya:2021twu,Horstmann:2022xkk}.\footnote{We mention that the authors of Ref.~\cite{Horstmann:2022xkk} did not directly extract the twist-3 function $g_{1T}^{(1)}(x)$ needed in our analysis.}  Previous numerical computations utilizing $g_{1T}^{(1)}(x)$ relied on a Wandzura-Wilczek approximation~\cite{Avakian:2007mv,Accardi:2009au,Kanazawa:2015ajw,Scimemi:2018mmi} that neglects quark-gluon-quark correlators to approximate $g_{1T}^{(1)}(x)$ in terms of an integral of the helicity PDF $g_{1}(x)$:~$g_{1T}^{(1)}(x) = x\int_x^1dy\,g_1(y)/y$.  In addition, the twist-3 fragmentation piece to $A_{LT}$ is sensitive to a coupling of the chiral-odd twist-3 FF $E(z)$ with $h_1(x)$~\cite{Koike:2015yza}.  No extractions exist of $E(z)$, but recent knowledge obtained about the closely related FF $\tilde{H}(z)$~\cite{Gamberg:2022kdb} allows us for the first time to develop a realistic input for $E(z)$ (in past numerical work, this function had been simply set to zero~\cite{Kanazawa:2014tda}).  The potential for future measurements of $A_{LT}$, particularly in electron-nucleon collisions, to provide more direct information about $E(z)$ are intriguing due to the connection of this FF to dynamical quark mass generation in QCD~\cite{Accardi:2017pmi,Accardi:2019luo,Accardi:2020iqn}.

From the experimental side, measurements of $A_{LT}$ in single-inclusive processes like those introduced above are unfortunately lacking.  The only data available are from Jefferson Lab 6 GeV (JLab6) on $A_{LT}$ in $\vec{e}\,n^\uparrow \to \pi\,X$~\cite{JeffersonLabHallA:2015vlz}.  Therefore, in this paper we give rigorous numerical predictions for $A_{LT}$ in a variety of reactions and kinematic configurations in order to motivate future measurements.  Namely, we will present results for $\vec{e}\,N^\uparrow \to \pi\,X$ for JLab 12 GeV (JLab12) with $N=n$, COMPASS with $N=p$, and the future Electron-Ion Collider (EIC) with $N=p$ (along with $\vec{e}\,p^\uparrow \to jet\,X$), as well as for the Relativistic Heavy Ion Collider (RHIC) for $p^\uparrow \vec{p}\to \{\pi, jet, \,{\rm or}\; \gamma\}\,X$.  Even with the new information about $g_{1T}^{(1)}(x)$ and $\tilde{H}(z)$ previously mentioned, we still must employ approximations for or neglect certain twist-3 PDFs or FFs due to lack of input for them. Thus, one stands to gain further insight into multi-parton correlations through measurements of $A_{LT}$.  Especially with only a few years of running left at RHIC, the world's only polarized proton-proton collider, one may forever lose the chance to measure $A_{LT}$ in $p^\uparrow \vec{p}\to \{\pi, jet,\;{\rm or}\; \gamma\}\,X$. 

The paper is organized as follows:~in Sec.~\ref{s:theory} we review the analytical formulas for $A_{LT}$ that have been derived in the literature for the processes of interest along with the twist-3 PDFs and FFs that enter them.  We also discuss the inputs and approximations used for these various non-perturbative functions as well as our strategy for computing the average values and uncertainties of our predictions.  We examine the main selected results for $A_{LT}$ in electron-nucleon and proton-proton collisions, and their implications for future measurements, in Sec.~\ref{s:results}.  The plots themselves can be can be found in Appendix~\ref{s:app_a} (for electron-nucleon) and Appendix~\ref{s:app_b} (for proton-proton).  In Sec.~\ref{s:concl} we close with our conclusions and outlook.

\section{Theoretical and Computational Background}
\label{s:theory}
In this section we review the analytical formulas for $A_{LT}$ needed for our computational work along with the relevant non-perturbative functions and certain relations between them.  The asymmetry itself is generically defined as
\begin{equation}
A_{LT} \equiv \frac{\dfrac{1}{4}\Big\{\!\left[d\sigma_{LT}(+,\uparrow) - d\sigma_{LT}(-,\uparrow)\right]-\left[d\sigma_{LT}(+,\downarrow) - d\sigma_{LT}(-,\downarrow)\right]\!\Big\}} {d\sigma_{unp}}\,, \label{e:ALT}
\end{equation}
where $d\sigma_{LT}(\lambda,\vec{S}_T)$ ($d\sigma_{unp}$) is the longitudinal-transverse spin-dependent (unpolarized) cross section, with $+$ ($-$) indicating a particle with positive (negative) helicity $\lambda$, and $\uparrow$ ($\downarrow$) denoting a particle with transverse spin $\vec{S}_T$ along the designated positive (negative) transverse axis (e.g., $\pm y$).  Moving forward, the numerator of Eq.~(\ref{e:ALT}) will be denoted by $d\sigma_{LT}$ (without any arguments).
We break this section down into the electron-nucleon and proton-proton cases.

\subsection{$\boldsymbol{A_{LT}}$ in Electron-Nucleon Collisions}
We consider the reaction $\vec{e}\,N^\uparrow\to \{\pi\;{\rm or}\;jet\}\,X$, where the produced final-state particle has a transverse momentum $P_T$, which sets the hard scale for the process.  We define the +z-axis to be the direction of $N^\uparrow$'s momentum in the electron-nucleon center-of-mass (c.m.) frame.  In addition to $P_T$, the asymmetry also depends on the c.m.~energy $\sqrt{S}$ and rapidity $\eta$ (which can also be written in terms of $x_F=2P_T\sinh(\eta)/\sqrt{S}$).  The coordinate system is such that at fixed-target experiments like JLab and COMPASS, the final-state particle is produced in the backward region (i.e., negative rapidity).  The two other Mandelstam variables at the hadronic level are 
$T= \left(-\sqrt{S}\,\sqrt{P_{T}^2 + x_F^2 S/4}+x_F S/2\right)$ and $U=\left(-\sqrt{S}\,\sqrt{P_{T}^2 + x_F^2S/4}-x_F S/2\right)$.
We can then write $A_{LT}$ for the case of pion production as~\cite{Kanazawa:2014tda,Kanazawa:2015ajw},
\begin{align}
A^{\vec{e}N^\uparrow\!\to\pi X}_{LT} = \frac{\displaystyle\int_{z_{min}}^1\dfrac{dz} {z^3}\!\left(\dfrac{-4P_{T}} {S+T/z}\right)\!\dfrac{1} {x}\displaystyle\sum_a e_a^2\left[\frac{M} {\hat{u}}\,D_1^{\pi/a}(z)\,\mathcal{G}^{a/N}\!(x,\hat{s},\hat{t},\hat{u}) + \dfrac{M_\pi} {z\hat{t}}\,h_1^{a/N}\!(x)\,E^{\pi/a}(z)\left(-\frac{\hat{s}}{\hat{t}}\right)\right]} {\displaystyle\int_{z_{min}}^1 \dfrac{dz} {z^2}\,\dfrac{1} {S+T/z}\,\dfrac{1} {x}\displaystyle\sum_a e_a^2\, f_1^{a/N}\!(x)\,D_1^{\pi/a}(z)\left(\dfrac{\hat{s}^2+\hat{u}^2} {\hat{t}^2}\right)}\,, \label{e:ALTeN} 
\end{align}
where
\begin{align}
\mathcal{G}(x,\hat{s},\hat{t},\hat{u})&= \left(g_{1T}^{(1)}\!(x)-x\frac{dg_{1T}^{(1)}\!(x)} {dx}\right)\!\left(\frac{\hat{s}(\hat{s}-\hat{u})} {2\hat{t}^{\hspace{0.025cm}2}}\right)+x\,g_T\!(x)\left(-\frac{\hat{s}\hat{u}}{\hat{t}^2}\right)+x\,g_1\!(x)\left(\frac{\hat{u}(\hat{s}-\hat{u})}{2\hat{t}^{\hspace{0.025cm}2}}\right), \label{e:scriptG}
\end{align}
 with $x= -(U/z)/(S+T/z)$, 
$z_{min} = -(T+U)/S$, and the partonic Mandelstam variables 
$\hat{s} = xS, \hat{t} = xT/z, \hat{u} = U/z$.  The sum $\sum_a$ is over all light quark and antiquark flavors ($a=q\;{\rm or}\; \bar{q}$), $e_a$ is the quark or antiquark charge (in units of the positron charge $e$), and $M$ ($M_\pi$) is the nucleon (pion) mass.  

The non-perturbative functions in Eqs.~(\ref{e:ALTeN}), (\ref{e:scriptG}) include the (twist-2) unpolarized PDF $f_1(x)$ and FF $D_1(z)$, helicity PDF $g_1(x)$, and transversity PDF $h_1(x)$, along with the kinematical twist-3 PDF $g_{1T}^{(1)}(x)$ (first-moment of the worm-gear TMD $g_{1T}$), intrinsic twist-3  PDF $g_T(x)$, and (chiral-odd) intrinsic twist-3 FF $E(z)$.  We see that Eq.~(\ref{e:ALTeN}) can be separated into two terms:~one involving twist-3 PDFs (what we will call the ``distribution term'') and one involving a twist-3 FF (what we will call the ``fragmentation term'').  We note that the case of jet production~\cite{Kang:2011jw} can be readily obtained from Eq.~(\ref{e:ALTeN}) by replacing $D_1(z)$ with $\delta(1-z)$ and setting the fragmentation term to zero.  

Some readers may be familiar with the more widely studied/measured $A_{LT}$ asymmetry in inclusive DIS $\vec{e}\,N^\uparrow\to e\,X$~\cite{Anthony:1996mw,Abe:1997qk,Abe:1998wq,Anthony:2002hy,Zheng:2004ce,Kramer:2005qe,Flay:2016wie,Armstrong:2018xgk}, where the scattered electron is detected in the final state instead of a pion.  In that process, the entire result depends only on $g_T(x)$, which is connected to the color Lorentz force on a struck quark in DIS~\cite{Burkardt:2008ps}. Already Eq.~(\ref{e:ALTeN}) makes apparent the rich structure of multi-parton correlators one is sensitive to in $A_{LT}$ for $\vec{e}\,N^\uparrow\to \pi\,X$ that cannot be accessed in inclusive DIS.  This presents both a challenge, in that one has several unknown twist-3 functions, but also an opportunity to probe different aspects of multi-parton correlations in hadrons.

As alluded to above, there are different categories of twist-3 correlators:~kinematical, intrinsic, and also dynamical~\cite{Kanazawa:2015ajw}.  The kinematical twist-3 functions are generically first-moments of twist-2 TMDs ($f^{(1)}(x)\equiv \int d^2\vec{k}_T \,\vec{k}_T^2/(2M^2)\,f(x,\vec{k}_T^2)$); intrinsic use a twist-3 Dirac projection in a quark-quark correlator; and dynamical are quark-gluon-quark or tri-gluon correlators.  These twist-3 PDFs or FFs are not independent of each other and can be related through QCD equation-of-motion relations (EOMRs) and Lorentz invariance relations (LIRs).  We refer the reader to Ref.~\cite{Kanazawa:2015ajw} (and references therein) for an extensive overview of collinear twist-3 functions, including their correlator definitions, derivations of EOMRs and LIRs, and how to express kinematical and intrinsic twist-3 functions in terms of the dynamical ones.  For the PDFs relevant to our study (see Eq.~(\ref{e:scriptG})), we note the following relations~\cite{Jaffe:1991ra,Tangerman:1994bb,Kotzinian:1995cz,Metz:2008ib,Accardi:2009au,Kanazawa:2015ajw}:
\begin{align}
    g_T^{q/N}\!(x) &= g_1^{q/N}\!(x)+\frac{dg_{1T}^{(1)q/N}\!(x)} {dx} - 2\mathcal{P}\!\int_{-1}^1 dy \,\frac{G_{FT}^{q/N}\!(x,y)}{(x-y)^2}\,,\label{e:LIR}\\[0.3cm]
    g_{1T}^{(1)q/N}(x) &= xg_T^{q/N}\!(x) -\frac{m_q}{M}\,h_1^{q/N}\!(x)+\mathcal{P}\int_{-1}^1dx_1 \frac{F_{FT}^{q/N}(x,x_1)-G_{FT}^{q/N}(x,x_1)}{x-x_1}\,,\label{e:EOMR}\\[0.3cm]
    g_T^{q/N}\!(x)&=\int_x^{\epsilon(x)} dy\,\frac{g_1^{q/N}\!(y)}{y} + \frac{m_q}{M}\left(\frac{h_1^{q/N}\!(x)}{x} + \int_{\epsilon(x)}^x dy\,\frac{h_1^{q/N}\!(y)}{y^2}\right)\nonumber\\ 
    &\hspace{0.2cm}+\,\int_x^{\epsilon(x)}\,\frac{dx_1}{x_1^2}\,\mathcal{P}\!\int_{-1}^1 dx_2\left[\frac{1-x_1\delta(x_1-x)}{x_1-x_2}F_{FT}^{q/N}\!(x_1,x_2) - \frac{3x_1-x_2-x_1(x_1-x_2)\delta(x_1-x)}{(x_1-x_2)^2}G_{FT}^{q/N}\!(x_1,x_2)\right]\,,\label{e:gTdyn}\\[0.3cm]
    g_{1T}^{(1)q/N}\!(x)&=x\int_x^{\epsilon(x)} dy\,\frac{g_1^{q/N}\!(y)}{y} + \frac{m_q}{M} x\int_{\epsilon(x)}^x dy\,\frac{h_1^{q/N}\!(y)}{y^2} \nonumber\\
    &\hspace{0.3cm} +\,x\int_x^{\epsilon(x)}\,\frac{dx_1}{x_1^2}\,\mathcal{P}\!\int_{-1}^1 dx_2\left[\frac{F_{FT}^{q/N}\!(x_1,x_2)}{x_1-x_2} - \frac{(3x_1-x_2)G_{FT}^{q/N}\!(x_1,x_2)}{(x_1-x_2)^2}\right]\,,\label{e:g1Tdyn}
\end{align}
where $\mathcal{P}$ denotes the principal value prescription, $\epsilon(x)\equiv 2\theta(x)-1$, $m_q$ is the quark mass, and $F_{FT}(x,x_1)$, $G_{FT}(x,x_1)$ are dynamical twist-3 PDFs (with $F_{FT}(x,x_1)$ giving the Qiu-Sterman function when $x=x_1$). The twist-2, kinematical twist-3, and intrinsic twist-3 PDFs all have support $-1\le x \le 1$, where $g_1^{q/N}\!(-x)=g_1^{\bar{q}/N}\!(x)$, $g_T^{q/N}\!(-x)=g_T^{\bar{q}/N}\!(x)$, $g_{1T}^{(1)q/N}\!(-x)=-g_{1T}^{(1)\bar{q}/N}\!(x)$, and $h_1^{q/N}\!(-x)=-h_1^{\bar{q}/N}\!(x)$.  The dynamical twist-3 PDFs have support $|x|\le 1$, $|x_1|\le 1$, and $|x-x_1|\le 1$, with $F_{FT}^{q/N}\!(-x_1,-x)=F_{FT}^{\bar{q}/N}\!(x,x_1)$ and $G_{FT}^{q/N}\!(-x_1,-x)=-G_{FT}^{\bar{q}/N}\!(x,x_1)$~\cite{Kanazawa:2015ajw}. The first expression~(\ref{e:LIR}) is a LIR and (\ref{e:EOMR}) is an EOMR, while (\ref{e:gTdyn}), (\ref{e:g1Tdyn}) are the result of solving Eqs.~(\ref{e:LIR}), (\ref{e:EOMR}) for the respective functions~\cite{Kanazawa:2015ajw} so that they only involve dynamical twist-3 correlators (with possibly a twist-2 term, as above with $\int_x^{\epsilon(x)} \!dy \,g_1(y)/y$).  Neglecting the quark mass terms and dynamical twist-3 PDFs in Eqs.~(\ref{e:gTdyn}), (\ref{e:g1Tdyn}) leads to the well-known Wandzura-Wilczek (WW) approximations~\cite{Wandzura:1977qf,Jaffe:1991ra,Tangerman:1994bb,Kotzinian:1995cz,Kotzinian:1997wt,Kotzinian:2006dw,Avakian:2007mv,Metz:2008ib,Accardi:2009au}
\begin{equation}
    g_T^{a/N}\!(x)\overset{{\rm WW}}{\approx}\int_x^1 dy\,\frac{g_1^{a/N}\!(y)}{y}\,,\quad\quad\quad
    g_{1T}^{(1)a/N}\!(x)\overset{{\rm WW}}{\approx}x\int_x^1 dy\,\frac{g_1^{a/N}\!(y)}{y} \,,\label{e:gTg1TWW}
\end{equation}
where $a=q\;{\rm or}\; \bar{q}$.  Until recently, the WW approximation was the only input available for $g_{1T}^{(1)}(x)$.  Now with the extraction of $g_{1T}^{(1)}(x)$ in Ref.~\cite{Bhattacharya:2021twu}, we do not necessarily have to resort to the WW approximation.  The expression in Eq.~(\ref{e:g1Tdyn}) makes clear there is more structure embedded in $g_{1T}^{(1)}(x)$ than what is accounted for in the WW approximation.  Likewise, using the extracted  $g_{1T}^{(1)}(x)$ from Ref.~\cite{Bhattacharya:2021twu} in Eq.~(\ref{e:LIR}) in principle inserts information about multi-parton correlators into the expression for $g_T(x)$, which the WW approximation does not encode.  Even so, we do not have complete information on $g_T(x)$ because $G_{FT}(x,x_1)$ is not known.  In Ref.~\cite{Bhattacharya:2020cen}, $g_{T}^{u-d}(x)$ was extracted for the first time in lattice QCD using the so-called quasi-distribution approach~\cite{Ji:2013dva}. An interesting prospect is one in principle could obtain information on $G_{FT}(x,x_1)$ through a flavor-separated computation of $g_T(x)$ on the lattice (taking $g_1(x)$ and $g_{1T}^{(1)}(x)$ as known functions).  

On the fragmentation side we have~\cite{Kanazawa:2015ajw}
\begin{align}
    E^{h/q}(z)=-2z\left(\int_z^\infty \frac{dz_1}{z_1^2}\,\frac{\hat{H}_{FU}^{\Re, h/q}(z,z_1)}{\frac{1}{z}-\frac{1}{z_1}}-\frac{m_q}{2M_h}D_1^{h/q}(z)\right), \label{e:E_EOMR}
\end{align}
where $\hat{H}_{FU}(z,z_1)$ is a quark-gluon-quark (dynamical twist-3) FF, and $M_h$ is the hadron mass.  The support properties are $0\le z \le 1$ and $z<z_1<\infty$~\cite{Kanazawa:2015ajw}. We mention again that dynamical twist-3 FFs are complex valued because of the lack of a time-reversal constraint in the fragmentation sector and have both real $\Re$ and imaginary $\Im$ parts. Recently, the FF $\tilde{H}(z)$ has been extracted~\cite{Gamberg:2022kdb}, and it is connected to the imaginary part of the same underlying correlator $\hat{H}_{FU}(z,z_1)$ as $E(z)$ depends on~\cite{Kanazawa:2015ajw}:
\begin{equation}
     \tilde{H}^{h/q}(z)=2z\int_z^\infty \frac{dz_1}{z_1^2}\,\frac{\hat{H}_{FU}^{\Im,h/q}(z,z_1)}{\frac{1}{z}-\frac{1}{z_1}}\,. \label{e:Htilde}
\end{equation}
We will use $\tilde{H}(z)$ to build up plausible scenarios for $E(z)$ in our numerical work.

\subsection{$\boldsymbol{A_{LT}}$ in Proton-Proton Collisions}

We now consider the reaction $p^\uparrow \vec{p}\to \{\pi, jet,\,{\rm or}\;\gamma\}\,X$.  We define the +z-axis to be the direction of $p^\uparrow$'s momentum in the proton-proton c.m.~frame.  There are three pieces to this observable for the case of pion production, depending on whether the twist-3 effects occur in $p^\uparrow$, $\vec{p}$, or $\pi$ (for $jet$ and $\gamma$, one only has the first two terms).  We write $A_{LT}$ for this case as
\begin{equation}
    A_{LT}^{p^\uparrow\vec{p}\to \pi X} = \frac{d\sigma_{LT}^{\rm Tdist}+d\sigma_{LT}^{\rm Ldist}+d\sigma_{LT}^{\rm frag}}{d\sigma_{unp}}\,,\label{e:ALTpp}
\end{equation}
where in the numerator we have indicated whether the term contains twist-3 effects from $p^\uparrow$ (transversely polarized distribution -- ``Tdist'')~\cite{Metz:2012fq}, from $\vec{p}$ (longitudinally polarized distribution -- ``Ldist'')~\cite{Koike:2016ura}, or from $\pi$ (fragmentation -- ``frag'')~\cite{Koike:2015yza}.  The expression for the unpolarized cross section reads
\begin{align}
d\sigma_{unp} &= \frac{\alpha_S^2}{S} \int_{z_{min}}^1 dz \int_{x_{min}}^1\frac{dx}{x}\frac{1}{x'z^2(xS+U/z)}\sum_i\sum_{a,b,c}\,f_1^{a/p}(x)\,f_1^{b/p}(x')\,D_1^{\pi/c}(z)\,H_U^i(\hat{s},\hat{t},\hat{u})\,,\label{e:sig_unp}
\end{align}
where $z_{min}=-(T+U)/S$, $x_{min}=-(U/z)/(S+T/z)$, $x' = -(xT/z)/(xS+U/z)$, and the summations are over all channels $i$ and parton flavors $a,b,c$.  The hard factors $H_U^i(\hat{s},\hat{t},\hat{u})$ depend on the partonic Mandelstam variables $\hat{s}=xx'S,\hat{t}=xT/z,\hat{u}=x'U/z$, and they can be found in Ref.~\cite{Kouvaris:2006zy}.

We next turn to the longitudinal-transverse polarized cross sections.  For $d\sigma_{LT}^{\rm Tdist}$ we have~\cite{Metz:2012fq}
\begin{align}
    d\sigma_{LT}^{\rm Tdist}  = -\frac{2\alpha_s^2MP_T}{S}\int_{z_{min}}^1 dz\int_{x_{min}}^1 \frac{dx}{x}\frac{1}{x'z^3(xS+U/z)}\sum_i\sum_{a,b,c}\frac{1}{\hat{m}_i}\,\mathcal{G}_i^{a/p^\uparrow}\!(x,\hat{s},\hat{t},\hat{u})\,g_1^{b/\vec{p}}(x')\,D_1^{\pi/c}(z)\,,\label{e:Tdist}
\end{align}
where
\begin{align}
\mathcal{G}_i(x,\hat{s},\hat{t},\hat{u}) &= \left(g_{1T}^{(1)}(x)-x\frac{dg_{1T}^{(1)}(x)}{dx}\right)H^i_{\tilde{g}}(\hat{s},\hat{t},\hat{u})
+ xg_T(x)\,H_{1,G_{DT}}^i(\hat{s},\hat{t},\hat{u}) +\frac{x}{2}\left(g_1(x)-g_T(x)\right)H_{3,G_{DT}}^i(\hat{s},\hat{t},\hat{u})\nonumber\\[0.1cm]
&\hspace{0.3cm}+\,\left[g_{1T}^{(1)}(x)+\mathcal{P}\int_{-1}^1 \frac{dx_1}{x_1}\,\frac{x\left(F_{FT}(x,x_1)+G_{FT}(x,x_1)\right)}{x-x_1}\right]H_{2,G_{DT}}^i(\hat{s},\hat{t},\hat{u})\,.\label{e:scriptGpp}
\end{align}
Some comments are in order about the expressions~(\ref{e:Tdist}), (\ref{e:scriptGpp}).  First, the variable $\hat{m}_i$ in Eq.~(\ref{e:Tdist}) is either $\hat{s}$, $\hat{t}$, or $\hat{u}$ depending on the channel $i$, with the specific values found in Table 1 of Ref.~\cite{Metz:2012fq}.\footnote{We note a typo in the last row for the $\hat{t}$ column of Table 1 in Ref.~\cite{Metz:2012fq}, where the channel should read $q\bar{q}\to \bar{q}'q'$.}  Second, the original expression in Ref.~\cite{Metz:2012fq} (see Eq.~(17) of that paper) is written in terms of the functions $\tilde{g}(x)$ and $F_{DT}(x,x_1), G_{DT}(x,x_1)$.  The former is just a different notation for $g_{1T}^{(1)}(x)$.  The latter are ``D-type'' dynamical twist-3 PDFs that use the covariant derivative, whereas we have chosen to write the result in terms of ``F-type'' functions $F_{FT}(x,x_1), G_{FT}(x,x_1)$ that use the field strength tensor. They are related via~\cite{Eguchi:2006qz}
\begin{align}
    F_{DT}(x,x_1) &= \mathcal{P}\frac{1}{x-x_1}\,F_{FT}(x,x_1)\,,\\
    G_{DT}(x,x_1) &= \mathcal{P}\frac{1}{x-x_1}\,G_{FT}(x,x_1) + \delta(x-x_1)\,g_{1T}^{(1)}(x)\,.
\end{align}
Lastly, we continued to ``optimize'' Eq.~(\ref{e:scriptGpp}) from the original version in Ref.~\cite{Metz:2012fq} so that it is written in terms of a maximal set of functions for which there is input for from the literature.  An observation made in Ref.~\cite{Metz:2012fq} was that the hard factors $H^i_{F_{DT}}$, $H^i_{G_{DT}}$ found in Appendix~A\footnote{The hard factors $H^i_{\tilde{g}}$ can also be found in Appendix A of Ref.~\cite{Metz:2012fq}.} of that paper can be broken down into three types of terms, namely, $H^i = H_1^i +H_2^i/(1-\xi)+H_3^i/\xi$, where $\xi=(x-x_1)/x$, with $H_{1,F_{DT}}^i=H^i_{1,G_{DT}}$, $H^i_{2,F_{DT}}=-H^i_{2,G_{DT}}$, and $H^i_{3,F_{DT}}=0$.  This insight allows one to use the LIR~(\ref{e:LIR}) and EOMR~(\ref{e:EOMR}) to obtain the final form in Eq.~(\ref{e:scriptGpp}), where now the only non-perturbative functions we lack input for are $F_{FT}(x,x_1),G_{FT}(x,x_1)$, and we will then ignore those terms in our numerical work.

We now give the formulas for the remaining two terms in the numerator of Eq.~(\ref{e:ALTpp}). For $d\sigma_{LT}^{\rm Ldist}$ we have~\cite{Koike:2016ura}
\begin{align}
    d\sigma_{LT}^{\rm Ldist} = -\frac{2\alpha_s^2MP_T}{S} \int_{z_{min}}^1 dz\int_{x_{min}}^1 \frac{dx}{x}\frac{1}{z^3(xS+U/z)}\sum_i\sum_{a,b,c}\,h_1^{a/p^\uparrow}\!\!(x)\,\mathcal{H}^{b/\vec{p}}(x',\hat{s},\hat{t},\hat{u})\,D_1^{\pi/c}(z)\,,\label{e:Ldist}
\end{align}
where
\begin{equation}
    \mathcal{H}(x',\hat{s},\hat{t},\hat{u}) = h_1(x')\,H_{1L}^i(\hat{s},\hat{t},\hat{u})+h_L(x')\,H_{2L}^i(\hat{s},\hat{t},\hat{u})+\frac{dh_{1L}^{\perp (1)}(x')}{dx'}\,H_{3L}^i(\hat{s},\hat{t},\hat{u})\,. \label{e:scriptH}
\end{equation}
The hard factors $H^i_{\{1,2,3\}L}$ correspond to $\hat{\sigma}_{\{1,2,3\}}$ in Eqs.~(16)--(21) of Ref.~\cite{Koike:2016ura}.  The function $h_L(x)$ is an intrinsic twist-3 function while $h_{1L}^{\perp(1)}(x)$ is kinematical twist-3 (first-moment of the other worm-gear TMD function $h_{1L}^{\perp}$).  Unlike $g_{1T}^{(1)}(x)$, there are no phenomenological extractions of $h_{1L}^{\perp(1)}(x)$.  Therefore, in our numerical work we must use WW approximations that connect $h_L(x)$ and  $h_{1L}^{\perp(1)}(x)$ to the twist-2 transversity PDF $h_1(x)$~\cite{Jaffe:1991ra,Tangerman:1994bb,Metz:2008ib,Kanazawa:2015ajw}:
\begin{equation}
    h_L^{a/N}\!(x)\overset{{\rm WW}}{\approx} 2x\int_x^1 dy\,\frac{h_1^{a/N}\!(y)}{y^2}\,,\quad\quad\quad
    h_{1L}^{\perp(1)a/N}\!(x)\overset{{\rm WW}}{\approx}x^2\int_x^1 dy\,\frac{h_1^{a/N}\!(y)}{y^2} \,,\label{e:hLh1LperpWW}
\end{equation}
where $a=q\;{\rm or}\; \bar{q}$.  Finally, for $d\sigma_{LT}^{\rm frag}$ we have~\cite{Koike:2015yza}
\begin{align}
    d\sigma_{LT}^{\rm frag} = \frac{2\alpha_s^2MP_T}{S} \int_{z_{min}}^1 dz\int_{x_{min}}^1 \frac{dx}{x}\frac{1}{x'z^4(xS+U/z)}\sum_i\sum_{a,b,c}\,h_1^{a/p^\uparrow}\!\!(x)\,g_1^{b/\vec{p}}(x')\,E^{\pi/c}(z)\,H_{f}^i(\hat{s},\hat{t},\hat{u})\,,\label{e:frag}
\end{align}
where the hard factors $H^i_{f}$ correspond to $\hat{\sigma}_i$ in Eq.~(15) of Ref.~\cite{Koike:2015yza}, and $E(z)$ is the same dynamical twist-3 FF introduced in the electron-nucleon case (\ref{e:ALTeN}) (see also Eq.~(\ref{e:E_EOMR})).

We mention that the result for $A_{LT}$ in $p^\uparrow \vec{p}\to jet \,X$ can be obtained by replacing $D_1(z)$ by $\delta(1-z)$ in Eqs.~(\ref{e:sig_unp}), (\ref{e:Tdist}), (\ref{e:Ldist}) and setting $d\sigma_{LT}^{\rm frag}$ to zero.  We refer the reader to Appendix B of Ref.~\cite{Metz:2012fq} (see also \cite{Liang:2012rb}) for the $d\sigma_{LT}^{\rm Tdist}$ formula for  $p^\uparrow \vec{p}\to \gamma \,X$.\footnote{Note that $\hat{m}_i=\hat{u}$ in this case for both channels ($qg\to \gamma q$ and $q\bar{q}\to \gamma g$), which was not explicitly stated in Ref.~\cite{Metz:2012fq}.}  To the best of our knowledge, the $d\sigma_{LT}^{\rm Ldist}$ formula for  $p^\uparrow \vec{p}\to \gamma \,X$ has not been derived yet in the literature.  Since we consider only direct photons, there is no $d\sigma_{LT}^{\rm frag}$ term.  The unpolarized cross section $d\sigma_{unp}$ for $pp\to \gamma \,X$ can be found in Ref.~\cite{Kouvaris:2006zy}.

\subsection{Numerical Methodology}\label{s:numerics}
We end this section with a discussion of our strategy for obtaining realistic numerical predictions for $A_{LT}$ given the information set forth in the previous two subsections.  

\subsubsection{Non-Perturbative Inputs}
With regard to input for the non-perturbative functions, we use CT18 NLO~\cite{Hou:2019qau} for $f_1(x)$, DSS14 NLO~\cite{deFlorian:2014xna} for $D_1(z)$, NNPDFpol1.1~\cite{Nocera:2014gqa} for $g_1(x)$, and JAM3D-22~\cite{Gamberg:2022kdb} for $h_1(x)$, all via LHAPDF 6.2.3~\cite{Buckley:2014ana}. For $g_{1T}^{(1)}(x)$ and $g_T(x)$ we consider two scenarios:
\begin{itemize}
    \item[(1)] {\bf quark-gluon-quark (qgq) scenario:}~We use $g_{1T}^{(1)}(x)$ extracted in Ref.~\cite{Bhattacharya:2021twu}, which in principle implicitly encodes dynamical twist-3 functions (see Eq.~(\ref{e:g1Tdyn})), and Eq.~(\ref{e:LIR}) for $g_T(x)$ with $G_{FT}(x,x_1)$ set to zero (since we have no direct input for it).  This is the maximal amount of information about quark-gluon-quark correlations we can include in $g_T(x)$ and $g_{1T}^{(1)}(x)$.
    \item[(2)] {\bf WW scenario:}~We use Eq.~(\ref{e:gTg1TWW}) for $g_T(x)$ and $g_{1T}^{(1)}(x)$, which completely neglects quark-gluon-quark correlations.
\end{itemize}

\begin{figure}[b!]
\includegraphics[width=0.75\textwidth]{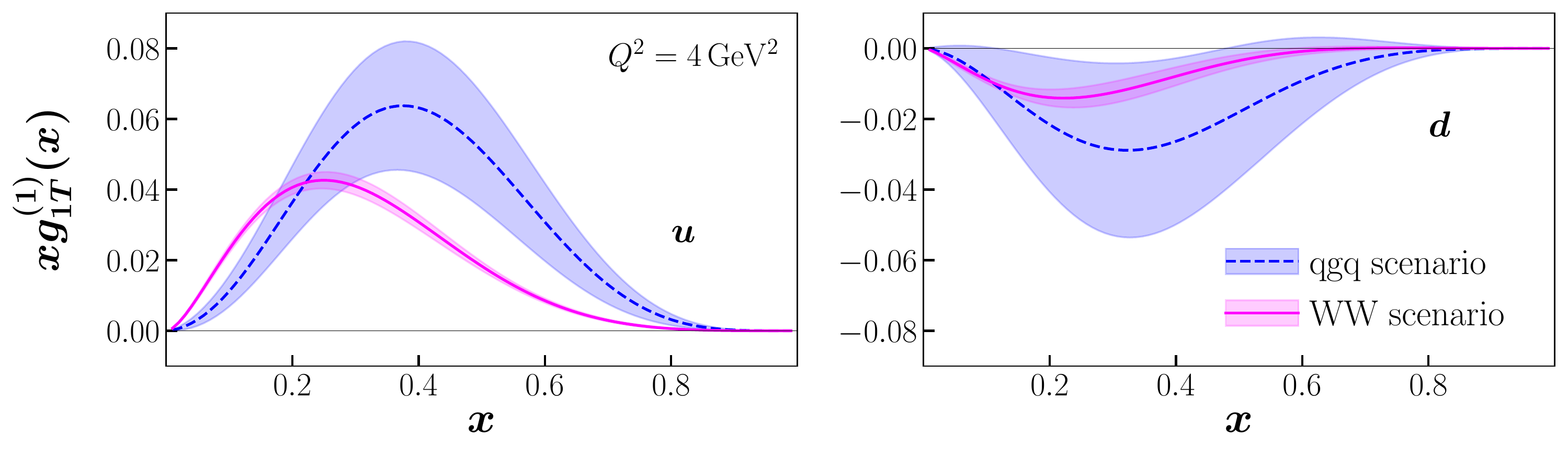}\vspace{-0.2cm}
\caption{Plot of the up ($u$) and down ($d$) quark in a proton kinematic twist-3 PDF $xg_{1T}^{(1)}(x)$ vs.~$x$ at $Q^2 = 4\,{\rm GeV^2}$ for the qgq scenario (blue dashed) and WW scenario (magenta solid) (both with 68\% C.L.~error bands). \vspace{-0.3cm}} 
\label{f:g1T_vs_x}
\end{figure}

\begin{figure}[t!]
\includegraphics[width=1\textwidth]{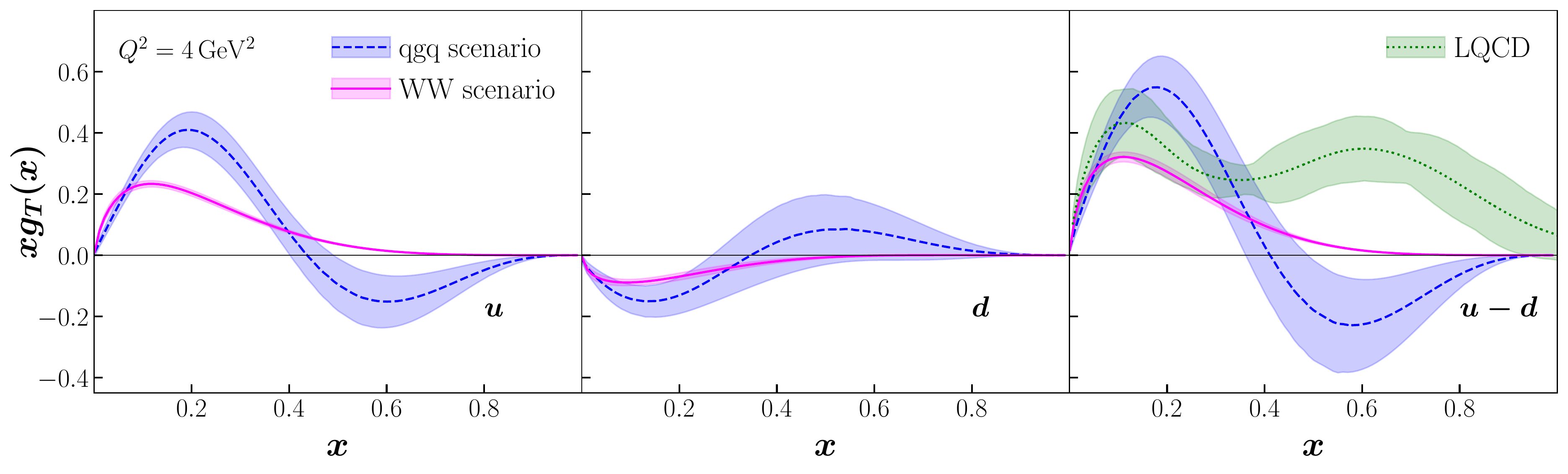}\vspace{-0.2cm}
\caption{Plot of the $u$, $d$, and $u-d$ in a proton intrinsic twist-3 PDF $xg_T(x)$ vs.~$x$ at $Q^2 = 4\,{\rm GeV^2}$ for the qgq scenario (blue dashed), WW scenario (magenta solid), and (for $u-d$) the lattice QCD (LQCD) calculation (green dotted) from Ref.~\cite{Bhattacharya:2020cen} (all with 68\% C.L.~error bands) . \vspace{-0.3cm}} 
\label{f:gT_vs_x}
\end{figure}

A plot comparing the two different scenarios for $g_{1T}^{(1)}(x)$ is shown in Fig.~\ref{f:g1T_vs_x}, and for $g_T(x)$ is shown in Fig.~\ref{f:gT_vs_x} along with a lattice QCD (LQCD) calculation (for the isovector $u-d$ combination) of the latter~\cite{Bhattacharya:2020cen}.\footnote{We note that the $g_T(x)$ computation in the qgq scenario depends on $g_1(x)$, where we use NNPDF replicas~\cite{Nocera:2014gqa}, and $g_{1T}^{(1)}(x)$, where we use the replicas from Bhattacharya, {\it et al.}~\cite{Bhattacharya:2021twu}.  To calculate the central curve and uncertainty band in this case, we use the same bootstrapping method described around Eq.~(\ref{e:Zstat}) below.}  We remark that $g_{1T}^{(1)u}(x)$ is larger in the qgq scenario and falls off slower at larger $x$.  Both the qgq and WW scenarios are compatible within error bands for $g_{1T}^{(1)d}(x)$.  The behavior of $g_T(x)$ in the two scenarios is quite different, mostly due to the $dg_{1T}^{(1)}(x)/dx$ term that enters Eq.~(\ref{e:LIR}) for the qgq case, which causes a change in sign in $g_T(x)$ at moderate $x$ values.  For the $d$ quark, the two scenarios are still compatible within error bands, but for the $u$ quark the qgq scenario is generally larger than the WW (in addition to having the aforementioned sign change).  The lattice computation for $g_T^{u-d}(x)$ shows agreement with the qgq and WW scenarios up to $x\approx 0.4$. At larger $x$, the WW scenario goes to zero the fastest, while the qgq scenario exhibits a change in sign and slower decrease as $x \to 1$.  The lattice calculation at large $x$ must deal with systematic effects in reconstructing the $x$ dependence that make the behavior of $g_T(x)$ in that region unreliable~\cite{Bhattacharya:2020cen}.  Once there is a rigorous lattice result of $g_T(x)$ across a wider range of $x$ and for individual $u$ and $d$ flavors, one in principle could use the difference between LQCD and the qgq scenario (taking $g_1(x)$ and $g_{1T}^{(1)}(x)$ as known functions) to extract information on the dynamical twist-3 PDF $G_{FT}(x,x_1)$ (see Eq.~(\ref{e:LIR})).

\begin{figure}[b!]
\includegraphics[width=0.75\textwidth]{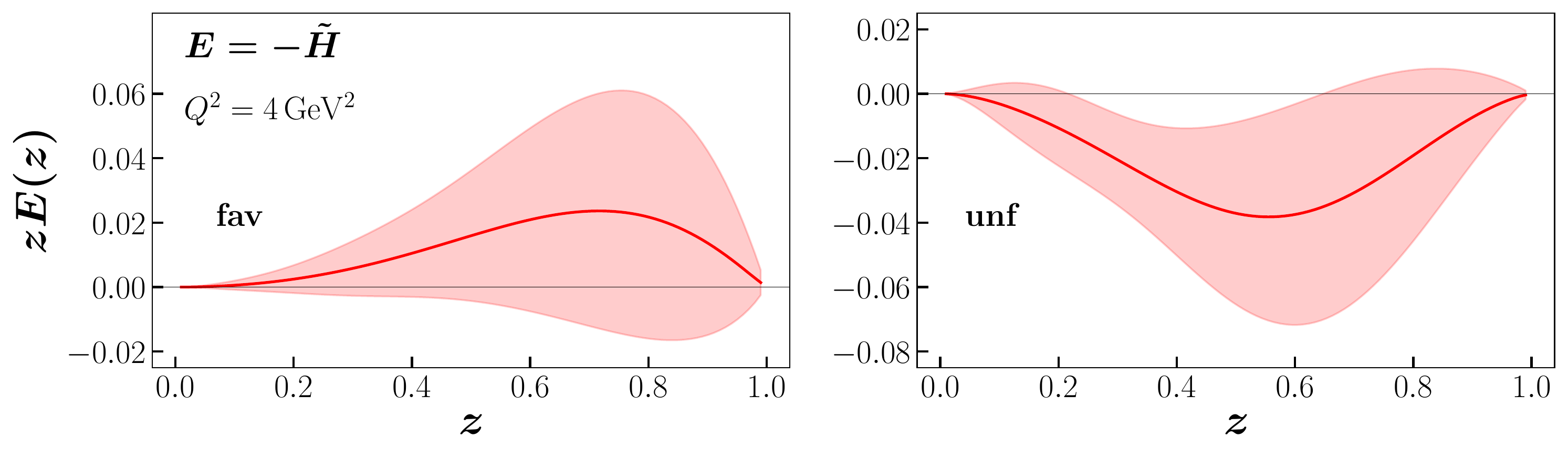}\vspace{-0.2cm}
\caption{Plot of the (favored and unfavored) twist-3 FF $zE(z)$ vs.~$z$ at $Q^2 = 4\,{\rm GeV^2}$ for the $E(z)=-\tilde{H}(z)$ scenario, where $\tilde{H}(z)$ is taken from Ref.~\cite{Gamberg:2022kdb}. \vspace{-0.3cm}} 
\label{f:E_vs_z}
\end{figure}

The last function we need input for is $E(z)$.  This intrinsic twist-3 FF was previously given attention in the literature because of its connection to dynamical quark mass generation in QCD~\cite{Accardi:2017pmi,Accardi:2019luo,Accardi:2020iqn}, which can also allow one to probe the transversity PDF $h_1(x)$ in {\it inclusive} DIS~\cite{Accardi:2017pmi}. As explicitly set forth in Eqs.~(\ref{e:E_EOMR}), (\ref{e:Htilde}), $E(z)$ is driven by the same quark-gluon-quark FF ($\hat{H}_{FU}(z,z_1)$) as $\tilde{H}(z)$, which we have input for from the JAM3D-22 analysis~\cite{Gamberg:2022kdb}.  Even so, there are some caveats with establishing this connection.  $E(z)$ depends on the real part of $\hat{H}_{FU}(z,z_1)$, while $\tilde{H}(z)$ depends on the imaginary part, and the two need not necessarily be related.  The functions also obey different sum rules~\cite{Accardi:2020iqn}:
\begin{equation}
    \sum_{h}\sum_{S_h} M_h\int_0^1 dz\,E^{h/q}(z)=M_j\,,\quad\quad\quad \sum_{h}\sum_{S_h} M_h\int_0^1 dz\,\tilde{H}^{h/q}(z)=0\,, \label{e:sumrules}
\end{equation}
where the summation is over all hadrons $h$ and their spins $S_h$.  The mass $M_j$ is the (gauge-invariant, non-perturbative) ``jet mass'' of a color-screened dressed quark propagating in the vacuum~\cite{Accardi:2019luo,Accardi:2020iqn}, which can be substantially larger than the current quark mass $m_q$.\footnote{The first term in Eq.~(\ref{e:E_EOMR}) can be identified as $\tilde{E}(z)$, which then allows for the decomposition $M_j=m_q+m_q^{corr}$ discussed in Refs.~\cite{Accardi:2019luo,Accardi:2020iqn}, where $M_j$ is broken down into the current quark mass $m_q$ and a term $m_q^{corr}$ that encodes dynamical mass generation due to quark-gluon-quark correlations.}  In the next section, we will revisit the possibility of $A_{LT}$ measurements, especially in electron-nucleon collisions, providing direct information about $E(z)$, and, therefore, potentially giving insight into $M_j$. These disclaimers notwithstanding, we think three realistic scenarios to study for $E(z)$ are $E(z)=-\tilde{H}(z)$, $E(z)=0$, and $E(z)=\tilde{H}(z)$. This accounts for $E(z)$ either being the same order of magnitude as $\tilde{H}(z)$ (although we cannot fix its sign) or $E(z)$ being significantly smaller than $\tilde{H}(z)$.  A plot for the $E(z)=-\tilde{H}(z)$ scenario is displayed in Fig.~\ref{f:E_vs_z}.

\subsubsection{Computation of Central Curves and Error Bands}
Clearly a numerical calculation of $A_{LT}$ in $\vec{e}\,N^\uparrow\to \{\pi\;{\rm or}\;jet\}\,X$ or $p^\uparrow \vec{p}\to \{\pi, jet,\,{\rm or}\;\gamma\}\,X$ depends on several non-perturbative inputs that have been extracted from various groups.  We now discuss our procedure for obtaining the central curves and error bands for the results presented in the next section. To aid in this explanation, we write the asymmetries as 
\begin{align}
    A^{\vec{e}N^\uparrow\!\to\pi X}_{LT} &=\frac{d\sigma_{LT}^{\rm dist}(g_1,g_{1T}^{(1)},g_T,D_1)+d\sigma_{LT}^{\rm frag}(h_1,E)}{d\sigma_{unp}(f_1,D_1)}\nonumber\\
    &\equiv A^{\vec{e}N^\uparrow\!\to\pi X}_{LT,{\rm dist}}(g_1,g_{1T}^{(1)},g_T,f_1,D_1)+A^{\vec{e}N^\uparrow\!\to\pi X}_{LT,{\rm frag}}(h_1,E,f_1,D_1)\,,\\[0.3cm]
    A_{LT}^{p^\uparrow\vec{p}\to \pi X} &= \frac{d\sigma_{LT}^{\rm Tdist}(g_1,g_{1T}^{(1)},g_T,D_1)+d\sigma_{LT}^{\rm Ldist}(h_1,D_1)+d\sigma_{LT}^{\rm frag}(h_1,g_1,E)}{d\sigma_{unp}(f_1,D_1)}\nonumber\\
    &\equiv A_{LT,{\rm Tdist}}^{p^\uparrow\vec{p}\to \pi X}(g_1,g_{1T}^{(1)},g_T,f_1,D_1) + A_{LT,{\rm Ldist}}^{p^\uparrow\vec{p}\to \pi X}(h_1,f_1,D_1)+A_{LT,{\rm frag}}^{p^\uparrow\vec{p}\to \pi X}(h_1,g_1,E,f_1,D_1)\,,\label{e:ALTppNP}
\end{align}
where we have explicitly indicated for each term which non-perturbative functions it depends on.\footnote{Note for $A_{LT,{\rm Ldist}}^{p^\uparrow\vec{p}\to \pi X}$, the non-perturbative functions that enter are $h_1(x)$, $h_L(x)$, and $h_{1L}^{\perp(1)}(x)$ (see Eqs.~(\ref{e:Ldist}), (\ref{e:scriptH})).  However, since we use WW approximations for the latter two, which depend on $h_1(x)$ (see Eq.~(\ref{e:hLh1LperpWW})), we have only denoted a dependence on $h_1(x)$.}  For $f_1(x)$ and $D_1(z)$, since they have relatively small uncertainties compared to the other PDFs and FFs, we simply use their central values and do not propagate their error into the computation.

We first focus on the electron-nucleon case. The fragmentation term $A^{\vec{e}N^\uparrow\!\to\pi X}_{LT,{\rm frag}}$ depends on $h_1(x)$ and $E(z)$ (recall we are using $\tilde{H}(z)$ to build our input for $E(z)$).  Both $h_1(x)$ and $\tilde{H}(z)$ were extracted simultaneously in JAM3D-22~\cite{Gamberg:2022kdb}, and we use all 450 replicas from that analysis to compute the mean and standard deviation for $A^{\vec{e}N^\uparrow\!\to\pi X}_{LT,{\rm frag}}$. For the distribution term, we are considering the two previously mentioned scenarios (WW and qgq).  In the WW scenario, $g_{1T}^{(1)}(x)$ and $g_T(x)$ both depend only on $g_1(x)$.  We therefore can use all 100 replicas from NNPDFpol1.1~\cite{Nocera:2014gqa} to determine the mean and standard deviation for $A^{\vec{e}N^\uparrow\!\to\pi X}_{LT,{\rm dist}}$.  The qgq scenario is more complicated because it depends on PDFs extracted by completely independent analyses, namely, $g_1(x)$ from NNPDFpol1.1~\cite{Nocera:2014gqa} and $g_{1T}^{(1)}(x)$ from Bhattacharya, {\it et al.}~\cite{Bhattacharya:2021twu} (recall our input for $g_T(x)$ depends on both these functions). For $g_{1T}^{(1)}(x)$ there are 200 replicas, so a complete calculation of $A^{\vec{e}N^\uparrow\!\to\pi X}_{LT,{\rm dist}}$ in the qgq scenario would require computing $100\times 200=20,\!000$ replicas.   Instead, we bootstrap the result by randomly sampling replicas for $g_1(x)$ and for $g_{1T}^{(1)}(x)$ (with replacement).  We continue to increase the number of replicas sampled and then calculate the (unequal variance or Welch's) $t$-statistic using the current and previous iterations, where~\cite{NumericalRecipes:2007} 
\begin{equation}
    t = \frac{\mu_1-\mu_2}{\sqrt{\sigma_1^2/N_1+\sigma_2^2/N_2}}\,, \label{e:Zstat}
\end{equation}
with $\mu$ the mean, $\sigma$ the standard deviation, and $N$ the number of ``data points'' (replicas sampled) of the respective distribution of $A_{LT}$ values for a given $P_T$.  Once $|t|$ is such that the corresponding $p$-values $\gtrsim 0.1$, then we consider the two distributions statistically equivalent~\cite{NumericalRecipes:2007} and do not proceed with any further iterations.\footnote{For many $P_T$ points, the $p$-values were much greater than 0.1, approaching 1.0 in some cases.}  (We also visually inspect the results to confirm the mean and standard deviation of $A_{LT}$ have converged.)
The $t$-statistic, and consequently the number of replicas required for convergence, is kinematic ($\sqrt{S},\,\eta,\,P_T$) and process (initial and final state) dependent.  For example, 1500 replicas were needed for JLab12 while 3000 were necessary for the EIC at $\sqrt{S}=29\,{\rm GeV}$.  Recall our calculation of $A^{\vec{e}N^\uparrow\!\to\pi X}_{LT,{\rm dist}}$ and $A^{\vec{e}N^\uparrow\!\to\pi X}_{LT,{\rm frag}}$ are totally uncorrelated from each other in that the respective non-perturbative functions that enter each term are from independent analyses by different groups.  Thus, once we have the final sample, we determine the central curve and uncertainty ($68\%$ C.L.~error band) as
\begin{equation}
\langle A^{\vec{e}N^\uparrow\!\to\pi X}_{LT}\rangle = \langle A^{\vec{e}N^\uparrow\!\to\pi X}_{LT,{\rm dist}}\rangle + \langle A^{\vec{e}N^\uparrow\!\to\pi X}_{LT,{\rm frag}} \rangle\,,  \quad\quad \delta \! A^{\vec{e}N^\uparrow\!\to\pi X}_{LT} = \sqrt{\left(\delta\! A^{\vec{e}N^\uparrow\!\to\pi X}_{LT,{\rm dist}}\right)^{\! 2} + \left(\delta\! A^{\vec{e}N^\uparrow\!\to\pi X}_{LT,{\rm frag}}\right)^{\! 2}}.\label{e:eN_avg_std}
\end{equation}

For the proton-proton case we follow a similar strategy, but there are some new aspects one must consider.  The fragmentation term $A_{LT,{\rm frag}}^{p^\uparrow\vec{p}\to \pi X}$ now also depends on $g_1(x)$ (since there is a longitudinally polarized proton involved, not an electron).  In addition, $A_{LT,{\rm Ldist}}^{p^\uparrow\vec{p}\to \pi X}$ depends on $h_1(x)$ and, consequently, must be computed simultaneously with $A_{LT,{\rm frag}}^{p^\uparrow\vec{p}\to \pi X}$ using the same replica sampled for $h_1(x)$ in that term.  Therefore, we must bootstrap the entire $A_{LT}^{p^\uparrow\vec{p}\to \pi X}$ asymmetry using the replicas from NNPDFpol1.1, Bhattacharya, {\it et al.}, and JAM3D-22, following a similar procedure as outlined for the electron-nucleon case, for both the WW and qgq scenarios.\footnote{Note that even for the WW scenario we need to employ bootstrapping since $g_1(x)$ shows up in $A_{LT,{\rm frag}}^{p^\uparrow\vec{p}\to \pi X}$.}  We again calculate the $t$-statistic of our $A_{LT}$ distributions (and visually inspect them) for different iterations to determine the number of samples required for convergence.  As before, there is a kinematic and process dependence; for example, RHIC $\sqrt{S}=200\, {\rm GeV}$ at midrapidity ($\eta=0$) needed 2500 samples while 3500 were necessary at forward rapidity ($\eta=3.3$).  Since all terms in $A_{LT}^{p^\uparrow\vec{p}\to \pi X}$ are correlated with each other, we determine the central curve and uncertainty using
\begin{equation}
\langle A^{p^\uparrow\vec{p}\to \pi X}_{LT}\rangle = \langle A^{p^\uparrow\vec{p}\to \pi X}_{LT,{\rm Tdist}}\rangle + \langle A_{LT,{\rm Ldist}}^{p^\uparrow\vec{p}\to \pi X} \rangle + \langle A_{LT,{\rm frag}}^{p^\uparrow\vec{p}\to \pi X} \rangle\,,  \quad\quad \delta \! A^{p^\uparrow\vec{p}\to \pi X}_{LT} = \delta\!\! \left(A^{p^\uparrow\vec{p}\to \pi X}_{LT,{\rm Tdist}}+A^{p^\uparrow\vec{p}\to \pi X}_{LT,{\rm Ldist}}+A^{p^\uparrow\vec{p}\to \pi X}_{LT,{\rm frag}}\right).
\end{equation}
We mention that for the jet and photon final states in proton-proton collisions, since the fragmentation term does not enter, the transverse and longitudinal distribution terms are uncorrelated.  The latter can be calculated using all replicas from JAM3D-22.  The former requires bootstrapping for the qgq scenario, but for the WW scenario it can be computed using all replicas from NNPDFpol1.1.  The central curve and uncertainty are then found exactly as in Eq.~(\ref{e:eN_avg_std}), with the replacements $(\vec{e}N^\uparrow\!\to\pi X) \longrightarrow (p^\uparrow\vec{p}\to \{jet\;{\rm or}\; \gamma\} X)$, ${\rm dist}\longrightarrow {\rm Tdist}$, ${\rm frag}\longrightarrow {\rm Ldist}$.

\section{Results and Discussion}
\label{s:results}
In this section we report our main results for $A_{LT}$ in electron-nucleon and proton-proton collisions.  We mention that, especially at the EIC and RHIC, we extensively studied the ($\sqrt{S},\,\eta,\,P_T$) coverage and are able to provide predictions for any reaction at any kinematics upon request.   Here we discuss a selective collection of plots, which can be found in Appendix~\ref{s:app_a} (for electron-nucleon) and Appendix~\ref{s:app_b} (for proton-proton), that highlight the main features of $A_{LT}$ in the single-inclusive processes under investigation.  Each plot shows six cases based on the possible combinations of input for $g_{1T}^{(1)}(x)$, $g_T(x)$, and $E(z)$, i.e., qgq or WW scenario for $g_{1T}^{(1)}(x)$, $g_T(x)$, and $E(z)=-\tilde{H}(z)$, $E(z)=0$, or $E(z)=\tilde{H}(z)$.  We remark again that the only measurement available of either $\vec{e}\,N^\uparrow \to \{\pi\;{\rm or}\; jet\}\,X$ or $p^\uparrow \vec{p}\to \{\pi, jet, \,{\rm or}\; \gamma\}\,X$ is from JLab6 for $\vec{e}\,n^\uparrow \to \pi\,X$~\cite{JeffersonLabHallA:2015vlz}.  There have been a few numerical calculations of $\vec{e}\,N^\uparrow \to \{\pi\;{\rm or}\; jet\}\,X$~\cite{Kang:2011jw,Kanazawa:2014tda}, but only with central curves (no error bands) using the WW approximation for $g_{1T}^{(1)}(x)$, $g_T(x)$ and (for pion production) ignoring the fragmentation term involving $E(z)$.  No numerical studies exist for the proton-proton case.

\subsection{Comparison with JLab6 Data} \label{s:JLab6}
 The comparison between our predictions and the JLab6 measurement is shown in Fig.~\ref{f:JLab6}. We caution that the data are at $P_T<1\,{\rm GeV}$, so one has to be careful about using a perturbative calculation in this region, and what conclusions to infer from it.  (In the computation, for any $P_T$-dependent kinematic quantities we used the actual experimental $P_T$ value, but in the non-perturbative functions we fixed $P_T=1\,{\rm GeV}$.)  We see that generally all cases are able to describe the data relatively well, with the distribution term playing a dominant role over the fragmentation term.  Nevertheless, there are hints, looking at the $E(z)=\tilde{H}(z)$ row of Fig.~\ref{f:JLab6}, that having a nonzero $E(z)$ with the same sign as $\tilde{H}(z)$ aids in obtaining better agreement with the data.  We note that the qgq scenario has larger error bands than the WW scenario because the direct extraction of $g_{1T}^{(1)}(x)$ is much less constrained than $g_1(x)$ (which is used in the WW approximation).  This is especially noticeable for $\pi^+$ because $g_{1T}^{d/p}(x)$ has a larger error band than $g_{1T}^{u/p}(x)$~\cite{Bhattacharya:2021twu} (recall JLab6  is for a neutron target, and we are employing isospin symmetry to obtain the neutron PDFs).

\subsection{Predictions for JLab12, COMPASS, and the EIC} \label{s:JL_COM_EIC}
We next give predictions for JLab12, COMPASS, and a few sets of EIC kinemtics.  We mention that next-to-leading order (NLO) corrections for the electron-nucleon single-inclusive  unpolarized  cross section ($eN\to \{\pi\,{\rm or}\, jet\}\,X$)~\cite{Hinderer:2015hra} have been shown to be sizeable, and for the double-longitudinal spin asymmetry $A_{LL}$ ($\vec{e}\,\vec{N}\to \{\pi\,{\rm or}\, jet\}\,X$)~\cite{Hinderer:2017ntk} they are also non-negligible. In addition, lower-energy experiments are  typically dominated by quasi-real photo-production~\cite{HERMES:2013quo}.  These issues should have less impact as one goes to higher $P_T$ ($\gtrsim 2\;{\rm or}\;3\,{\rm GeV}$), but high-precision measurements at the EIC may require NLO calculations.  

In Fig.~\ref{f:JLab12} we present results for JLab12 with a neutron target.  In all cases, sizeable asymmetries $\sim \!15$-$30\%$ are predicted which grow more substantial with increasing $P_T$.  The distribution term gives basically the entirety of $A_{LT}$.  The qgq scenario also tends to be larger than the WW scenario, especially at higher $P_T$.  Therefore, one may be able to use JLab12 data to test the WW approximation and potentially extract information about dynamical quark-gluon-quark correlations in the nucleon.  

The COMPASS results are displayed in Fig.~\ref{f:COMPASS} for a proton target, which are roughly an order of magnitude smaller than JLab12 but still measurable at $\sim \!2$-$4\%$.  From the first ($E(z)=-\tilde{H}(z)$) and last  ($E(z)=\tilde{H}(z)$) rows of the plot, we see that, unlike JLab12, the $A_{LT}$ fragmentation term can be comparable to the distribution term, at least for $\pi^-$ production.  Since the $E(z)=0$  case  (middle row) has $A_{LT}$ for $\pi^-$ clearly positive, a measured negative asymmetry would be a likely indication of quark-gluon-quark fragmentation effects.  The qgq and WW scenarios may be difficult to distinguish at COMPASS since they give similarly-sized effects.

The low-energy EIC predictions at midrapidity ($\sqrt{S}=29\,{\rm GeV},\,\eta=0$) are shown in Fig.~\ref{f:EIC29_eta0}, where again we notice a further decrease in the size of the asymmetry compared to JLab12 and COMPASS, with $A_{LT}$ now $\sim \!0.5$-$1.5\%$.  Similar to COMPASS, a clearly negative signal for $\pi^-$ production would be caused by quark-gluon-quark fragmentation.  Since the EIC will also measure jets, we give results for that reaction at higher-energy EIC kinematics and slightly forward rapidity ($\sqrt{S}=63\,{\rm GeV},\,\eta=1$) in Fig.~\ref{f:EIC63jet_eta1}.  The asymmetry again decreases, now to $\sim \!0.1$-$0.3\%$, due to the increase in c.m.~energy and the fact that jets are being detected instead of pions.

The general features of $A_{LT}$ in electron-nucleon collisions are that it increases with $P_T$ but decreases significantly with $\sqrt{S}$.  However, as $\eta$ increases, and one pushes $P_T$ to the theoretical kinematic limit, the fragmentation term can cause an enhanced growth in $A_{LT}$.  A typical example is shown in Fig.~\ref{f:EIC29_eta1}. One sees the asymmetry is basically zero for most of the $P_T$ range and then receives an sizeable enhancement at the largest $P_T$ values.  In this region, $z_{min}$ in Eq.~(\ref{e:ALTeN}) is around $0.8$ to $0.9$; one is then integrating at the threshold of producing the pion, where $E(z)$ is not constrained and resummation techniques may be needed~\cite{Anderle:2012rq,Anderle:2013lka,Hinderer:2014qta,Hinderer:2018nkb,Kaufmann:2019ksh}.  Whether or not this is a physical effect that would be observed in experiments remains to be seen.

The measurement of $A_{LT}$ in $\vec{e}\,N^\uparrow \to \{\pi\;{\rm or}\; jet\}\,X$ at future experiments has the potential to provide insight into quark-gluon-quark correlations, especially given the precision expected at the EIC.  A reduction in the uncertainty of $g_{1T}^{(1)}(x)$ will be key if one is to disentangle dynamical twist-3 effects from the twist-2 WW approximation.  More precise measurements of the $A_{LT}^{\cos(\phi_h-\phi_S)}$ modulation in SIDIS at COMPASS, SoLID at JLab, and the EIC   will be crucial to achieve this.  For example, there are hints in Fig.~\ref{f:EIC29_eta0} that the qgq scenario may differ from the WW scenario by $\sim0.5\%$, but currently the error band in the qgq scenario (that relies on the full extraction of $g_{1T}^{(1)}(x)$) is too large to distinguish the two.  A similar statement can be made for jet production in Fig.~\ref{f:EIC63jet_eta1}. Also recall that even in the qgq scenario, we neglected the dynamical twist-3 PDF $G_{FT}(x,x_1)$ in Eq.~(\ref{e:LIR}).  Thus, significant differences between the qgq scenario predictions and future data could provide information on this function. Moreover, any significant deviations from the $E(z)=0$ scenario, especially if $g_{1T}^{(1)}(x)$ becomes more constrained, would allow for an extraction of this twist-3 FF.  Given its connection to dynamical quark mass generation in QCD (see the discussion around Eq.~(\ref{e:sumrules})), the potential for $A_{LT}$ to give us information on $E(z)$ is another intriguing reason to measure it.

\subsection{Predictions for RHIC} \label{s:RHIC}

We now report on the results for $A_{LT}$ in $p^\uparrow \vec{p}\to \{\pi, jet, \,{\rm or}\; \gamma\}\,X$ at RHIC, the only machine capable of measuring this asymmetry.  We focus on $\sqrt{S}=200\,{\rm GeV}$ c.m.~energy at middle and forward rapidities.  We remind the reader that there are three pieces to the asymmetry given in Eqs.~(\ref{e:Tdist}), (\ref{e:Ldist}), (\ref{e:frag}) (although the fragmentation term doesn't enter for photon or jet production).  Our predictions for charged pion production at midrapidity ($\eta=0$) in Fig.~\ref{f:RHIC_mid_pipm} reach to $\sim \!0.02$-$0.05\%$ for $\pi^\pm$ at the highest $P_T$.  The transverse distribution term gives the largest contribution to $A_{LT}$, although the fragmentation term plays a non-negligible role.  At forward rapidity ($\eta=3.3$) in Fig.~\ref{f:RHIC_forward_pipm}, the asymmetry has larger error bands for the qgq scenario that are consistent with zero but range from $\sim \!-0.3\%$ to $+0.2\%$.  In the WW approximation the uncertainties are much smaller at larger $P_T$ and again consistent with zero.  In either case, the transverse distribution term gives the entirety of $A_{LT}$ at forward rapidity.   The $\pi^0$ asymmetries (Figs.~\ref{f:RHIC_mid_pi0}, \ref{f:RHIC_forward_pi0}) are similar in size to $\pi^\pm$.  For jet or photon production at midrapidity (Fig.~\ref{f:RHIC_mid_jet_gam}), our predictions for $A_{LT}$ are $\lesssim 0.03\%$. We note that at $\sqrt{S}=500\,{\rm GeV}$, the asymmetry (for any final state) is generally an order of magnitude smaller than at $\sqrt{S}=200\,{\rm GeV}$.

The reader may question why $A_{LT}$ in proton-proton collisions is much smaller than $A_N$.  Recall that $A_N$ (where one proton is unpolarized and the other is transversely polarized) is another (much more widely studied/measured) twist-3 asymmetry that {\it does} show significant effects, at least in the forward region~\cite{Adams:1991rw,Krueger:1998hz,Allgower:2002qi,Adams:2003fx,Adler:2005in,Lee:2007zzh,Abelev:2008af,Arsene:2008aa,Adamczyk:2012qj,Adamczyk:2012xd,Bland:2013pkt,Adare:2013ekj,Adare:2014qzo,STAR:2020nnl}. We found that there are two driving factors.  First, in the $qg\to qg$ channel (which is the dominant channel in the numerator of $A_N$ and $A_{LT}$), the fragmentation term for $A_N$ (which is the main source of the asymmetry~\cite{Kanazawa:2014dca,Gamberg:2017gle,Cammarota:2020qcw,Gamberg:2022kdb}) has hard factors $\sim 1/\hat{t}^3$, whereas in the transverse distribution term (\ref{e:Tdist}) for $A_{LT}$ (which is the main source of that asymmetry) the hard factors $\sim 1/(\hat{t}^2\hat{u})$.  Since $\hat{t}\to 0$ in the forward region, this provides an enhancement to $A_N$ not seen in $A_{LT}$. The second difference is $A_N$ has an unpolarized proton, so in the $qg\to qg$ channel, $f_1^g(x)$ multiplies the (twist-3) fragmentation term.  On  the other hand, $A_{LT}$ has a longitudinally polarized proton, so $g_1^g(x)$ multiplies the (twist-3) transverse distribution term.  In the forward region (of the transversely polarized proton), these gluon functions are probed at small $x$; hence, $A_N$ becomes signficantly larger than $A_{LT}$.  In fact, we checked that if in the numerator of $A_N$ one replaces $f_1^g(x)$ (in the $qg\to qg$ channel) with $g_1^g(x)$, the asymmetry is nearly as suppressed as $A_{LT}$.

We emphasize that, in addition to the assumptions that underlie our scenarios for $g_{1T}^{(1)}(x),g_T(x)$ and $E(z)$, the proton-proton case has several terms that we are forced to neglect due to lack of input for dynamical twist-3 correlators.  Namely, we do not consider the terms in Eq.~(\ref{e:scriptGpp}) involving $F_{FT}(x,x_1), G_{FT}(x,x_1)$.  The WW approximation we use for $h_L(x)$ and $h_{1L}^{\perp(1)}(x)$ in Eq.~(\ref{e:scriptH}) sets to zero another dynamical twist-3 PDF called $H_{FL}(x,x_1)$~\cite{Jaffe:1991ra,Tangerman:1994bb,Metz:2008ib,Kanazawa:2015ajw}.\footnote{We note that there are some model calculations of functions connected to $F_{FT}(x,x_1), G_{FT}(x,x_1)$~\cite{Braun:2011aw}. The worm-gear TMD $h_{1L}^{\perp}$ in the future can be extracted from data on the $A_{LT}^{\sin 2\phi_h}$ modulation in SIDIS~\cite{HERMES:1999ryv,HERMES:2001hbj,HERMES:2002buj,CLAS:2010fns,COMPASS:2016klq,HERMES:2020ifk}.}  Therefore, measurements that significantly deviate from our predictions could provide information on these unknown quark-gluon-quark correlators. 

\section{Conclusions and Outlook}
\label{s:concl}
We have numerically analyzed the twist-3 asymmetry $A_{LT}$ in single-inclusive electron-nucleon and proton-proton collisions for various final states.  This is the first time  contributions from all terms entering these asymmetries have been computed. Nevertheless,  some approximations/assumptions had to be employed, including ignoring certain dynamical twist-3 PDFs due to a lack of information about them.  Using recent extractions of $g^{(1)}_{1T}(x)$~\cite{Bhattacharya:2021twu} and $\tilde{H}(z)$~\cite{Gamberg:2022kdb}, we were able to develop realistic scenarios to investigate for three critical functions in $A_{LT}$:~$g_{1T}^{(1)}(x)$, $g_T(x)$, and $E(z)$.  We used bootstrapping to provide a rigorous error quantification of our calculation that accounts for the fact that $A_{LT}$ depends on multiple non-perturbative functions extracted by different groups.  We found good agreement with JLab6 data, which is the only $A_{LT}$ measurement available (for single-inclusive observables).  We then made predictions for $A_{LT}$ in electron-nucleon collisions at JLab12, COMPASS, and the  EIC, as well as proton-proton collisions at RHIC, in order to motivate future measurements. Beyond the results presented in this paper, we are able to provide predictions for any initial/final states and kinematic region $(\sqrt{S},\eta,P_T)$ upon request.

In electron-nucleon collisions, the asymmetry decreases with increasing center-of-mass energy, going from (for $\pi^\pm$ production) $\sim \!15$-$30\%$ at JLab12 to $\sim \!2$-$4\%$ at COMPASS to $\sim \!0.5$-$1.5\%$ for the low-energy EIC configuration (at midrapidity). An  intriguing prospect is if significant deviations from the $E(z)=0$ scenario are measured, it could provide direct information on $E(z)$, which is connected to dynamical quark mass generation in QCD~\cite{Accardi:2017pmi,Accardi:2019luo,Accardi:2020iqn}.  One may also be able to test the validity of the Wandzura-Wilczek approximation for $g^{(1)}_{1T}(x),g_T(x)$ and probe dynamical twist-3 PDFs, especially with precision measurements at the EIC.  The calculation of the proton-proton case at RHIC kinematics  showed (for $\pi^\pm$ production) $A_{LT}\sim \!0.02$-$0.05\%$ at midrapidity and can be in the range of $\sim \!-0.3\%$ to $+0.2\%$ at forward rapidity. The asymmetry does not grow rapidly at forward rapidity, in contrast to $A_N$, due to a suppression caused by the other proton being longitudinally polarized instead of unpolarized (where $g_1^g(x)$ then enters the $qg\to qg$ channel in the numerator of the asymmetry instead of $f_1^g(x)$).  Since RHIC is the only machine capable of measuring $A_{LT}$ in proton-proton collisions, confirmation or refutation of our predictions would aid in better understanding the role of quark-gluon-quark correlations in hadrons.

\section*{Acknowledgments}  
This work has been supported by the National Science Foundation under Grant No.~PHY-2011763.  The authors thank S.~Bhattacharya for providing the lattice data of Ref.~\cite{Bhattacharya:2020cen} and for valuable feedback from a careful reading of the manuscript.  The authors also thank E.~Aschenauer, A.~Metz, N.~Sato, and R.~Seidl for fruitful discussions about various aspects of this work.  The authors are also grateful to C.~Cocuzza for creating LHAPDF tables of the JAM3D-22 functions, to R.~Abdul Khalek for providing the LHAPDF tables of DSS14 created by V.~Bertone, and to Jefferson Lab for access to their computational resources.

\appendix
\section{Electron-Nucleon Results}
\label{s:app_a}
\begin{figure}[h!]
\includegraphics[width=0.625\textwidth]{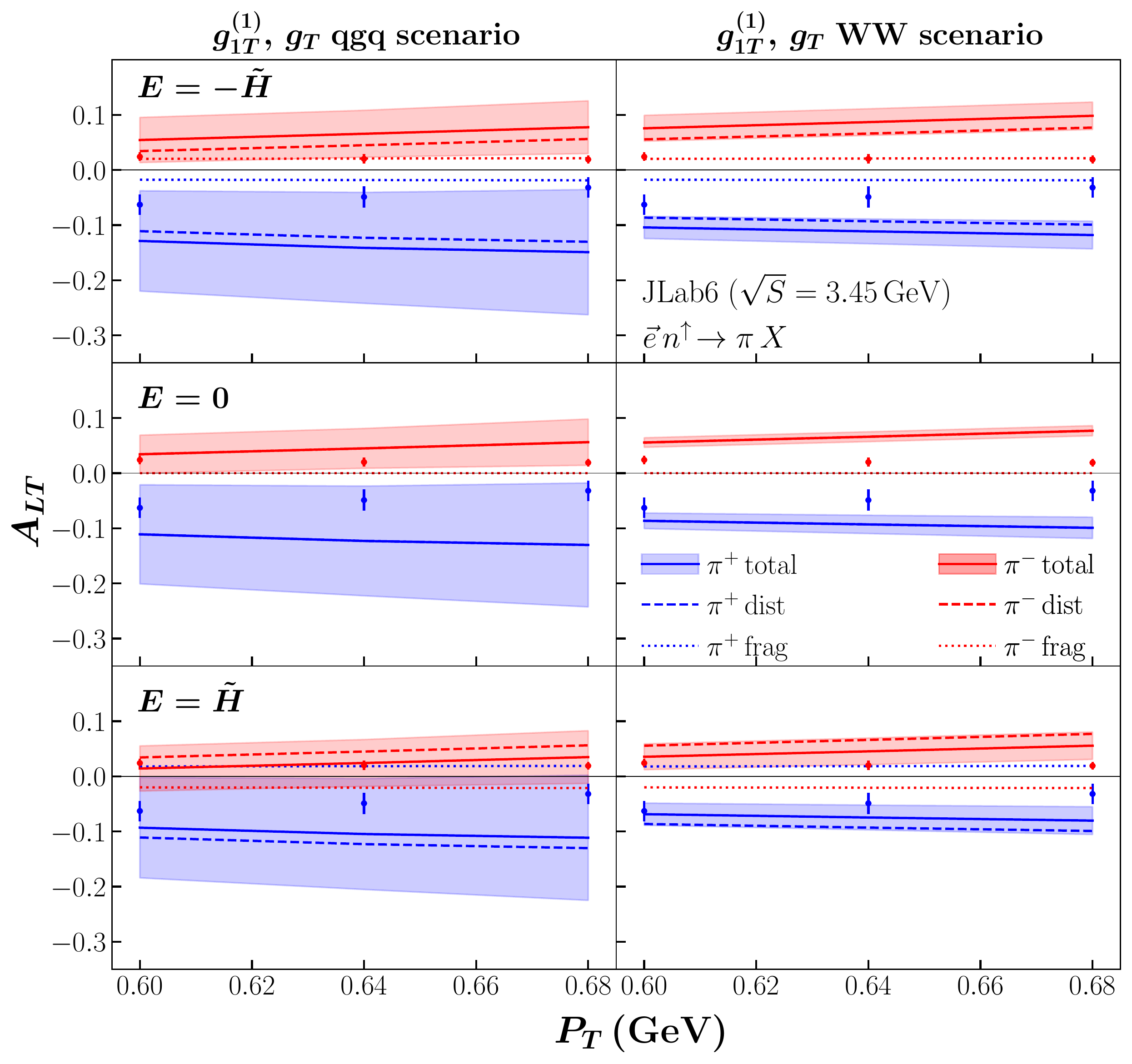}\vspace{-0.2cm}
\caption{Predictions for $A_{LT}$ vs.~$P_T$ in $\vec{e}\,n^\uparrow\!\to \pi\,X$ compared to JLab6 data~\cite{JeffersonLabHallA:2015vlz}.  The left column is for the qgq scenario for $g_{1T}^{(1)}(x)$, $g_T(x)$ and the right is for the WW scenario (see Sec.~\ref{s:numerics} for more details).  The first row is for the case $E(z)=-\tilde{H}(z)$, the second for $E(z)=0$, and third for $E(z)=\tilde{H}(z)$.  The solid curve gives the average total asymmetry (with $68\%$ C.L.~error band), while the dashed (dotted) curves give the average individual contribution from the distribution (fragmentation) term. \vspace{-0.3cm}} 
\label{f:JLab6}
\end{figure}

\begin{figure}[h!]
\includegraphics[width=0.625\textwidth]{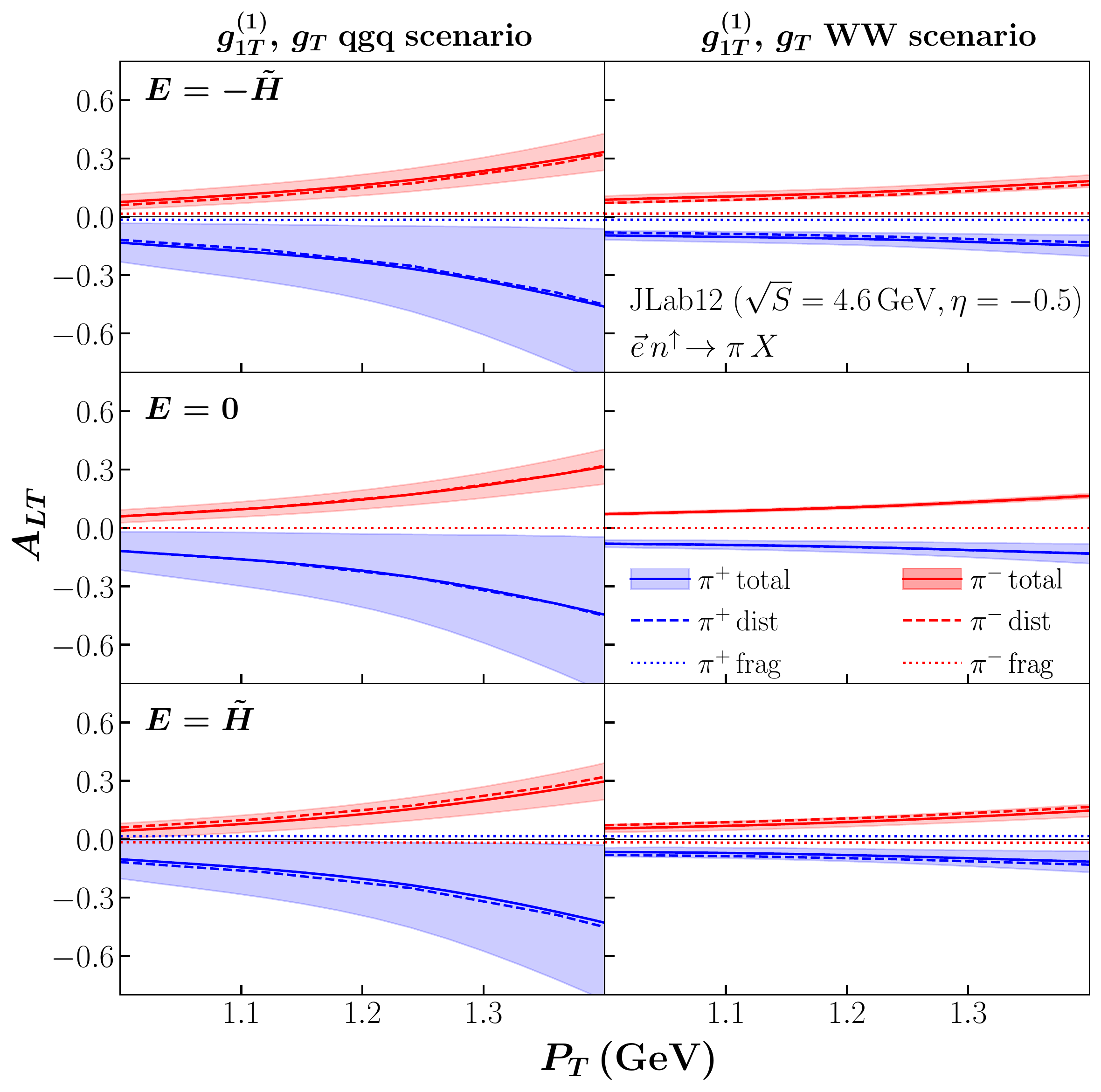}\vspace{-0.2cm}
\caption{Predictions for $A_{LT}$ vs.~$P_T$ in $\vec{e}\,n^\uparrow\!\to \pi\,X$ for JLab12 kinematics ($\sqrt{S}=4.6\,{\rm GeV},\eta=-0.5$).  The description is the same as the Fig.~\ref{f:JLab6} caption.\vspace{-0.3cm}} 
\label{f:JLab12}
\end{figure}

\begin{figure}[h!]
\includegraphics[width=0.625\textwidth]{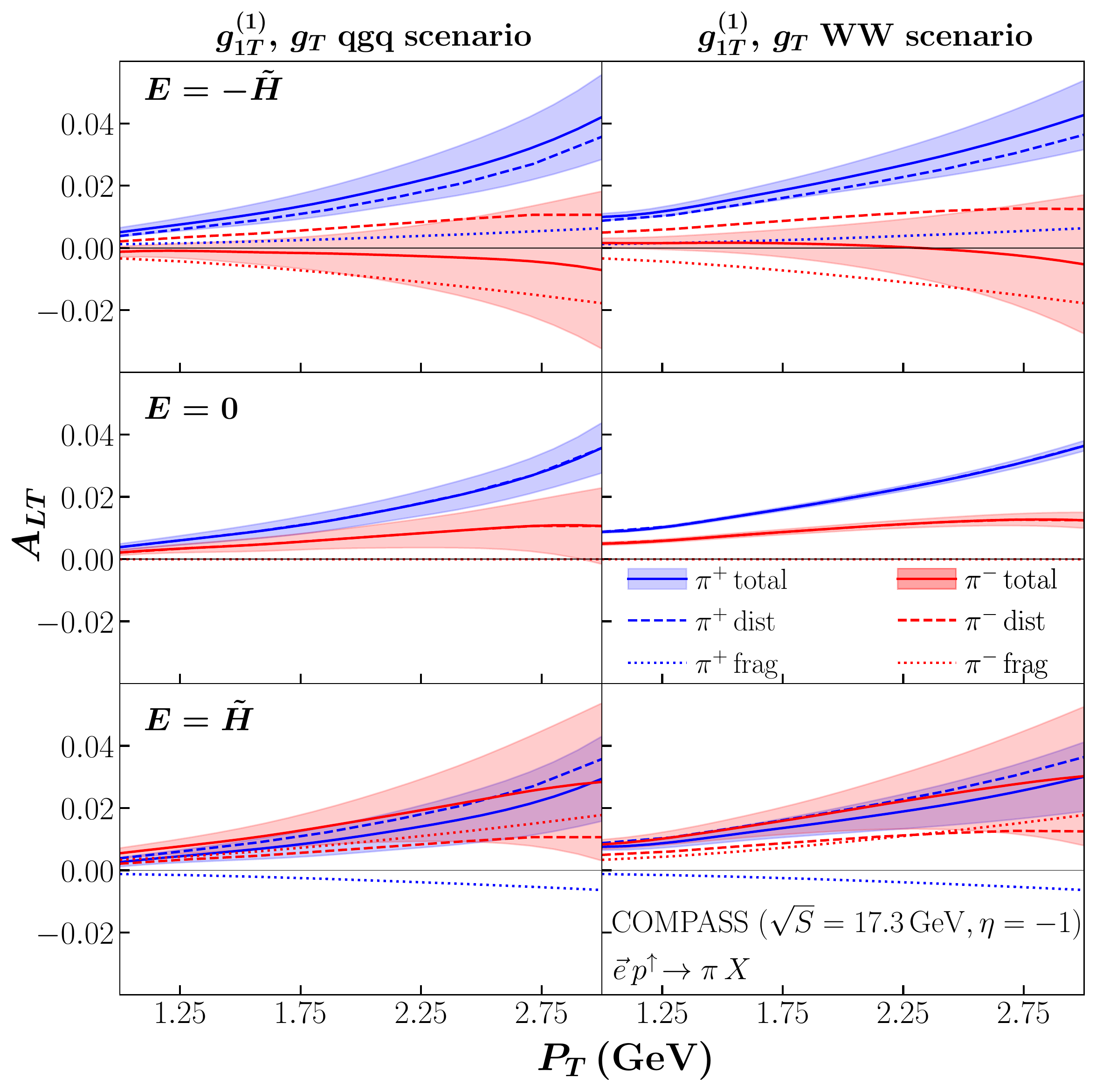}\vspace{-0.2cm}
\caption{Predictions for $A_{LT}$ vs.~$P_T$ in $\vec{e}\,p^\uparrow\!\to \pi\,X$ for COMPASS kinematics ($\sqrt{S}=17.3\,{\rm GeV},\eta=-1$).   The description is the same as the Fig.~\ref{f:JLab6} caption.\vspace{-0.7cm}} 
\label{f:COMPASS}
\end{figure}

\begin{figure}[h!]
\includegraphics[width=0.625\textwidth]{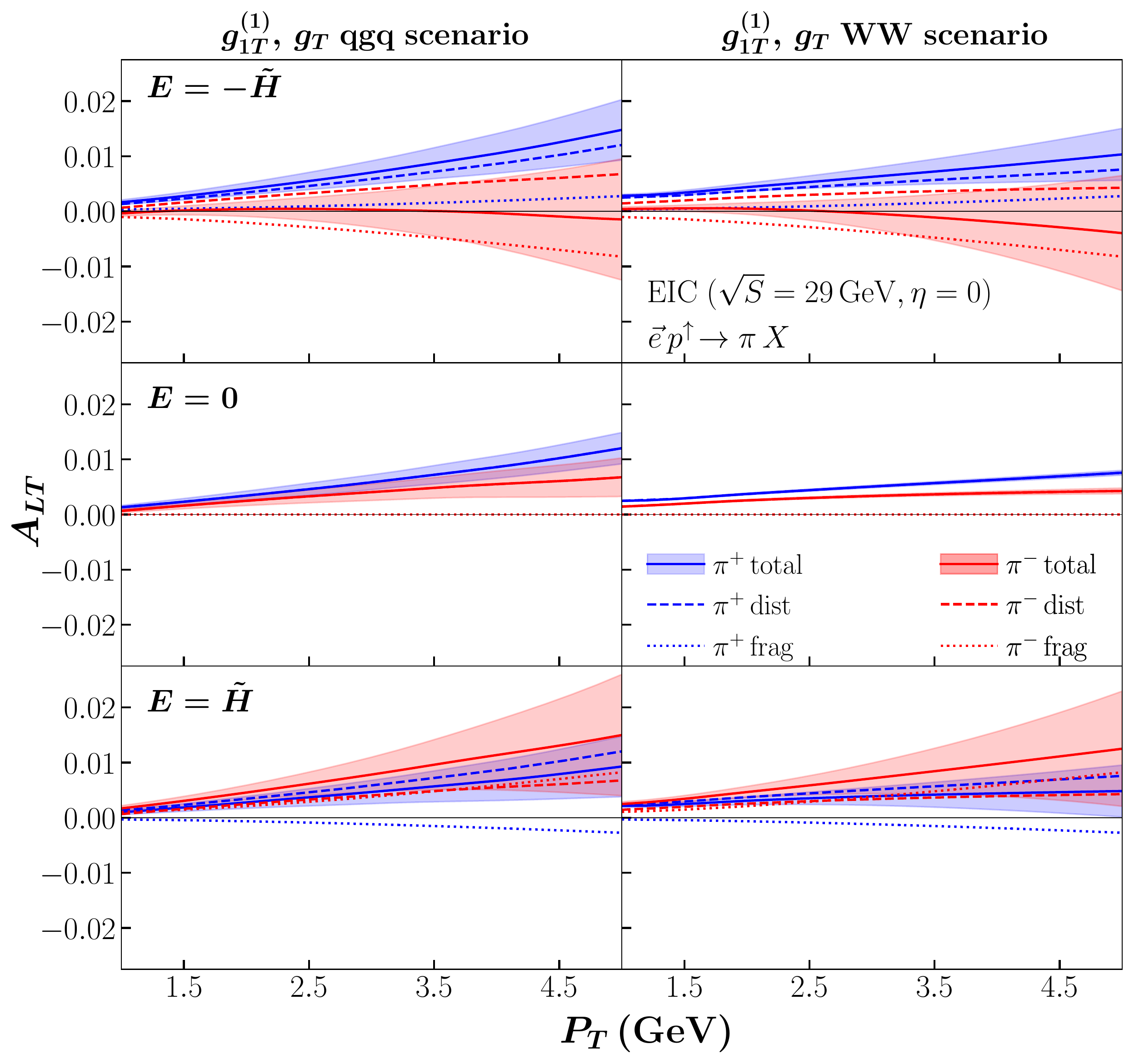}\vspace{-0.2cm}
\caption{Predictions for $A_{LT}$ vs.~$P_T$ in $\vec{e}\,p^\uparrow\!\to \pi\,X$ for low-energy EIC kinematics at midrapidity ($\sqrt{S}=29\,{\rm GeV},\eta=0$).   The description is the same as the Fig.~\ref{f:JLab6} caption.\vspace{-0.3cm}} 
\label{f:EIC29_eta0}
\end{figure}

\begin{figure}[h!]
\includegraphics[width=0.625\textwidth]{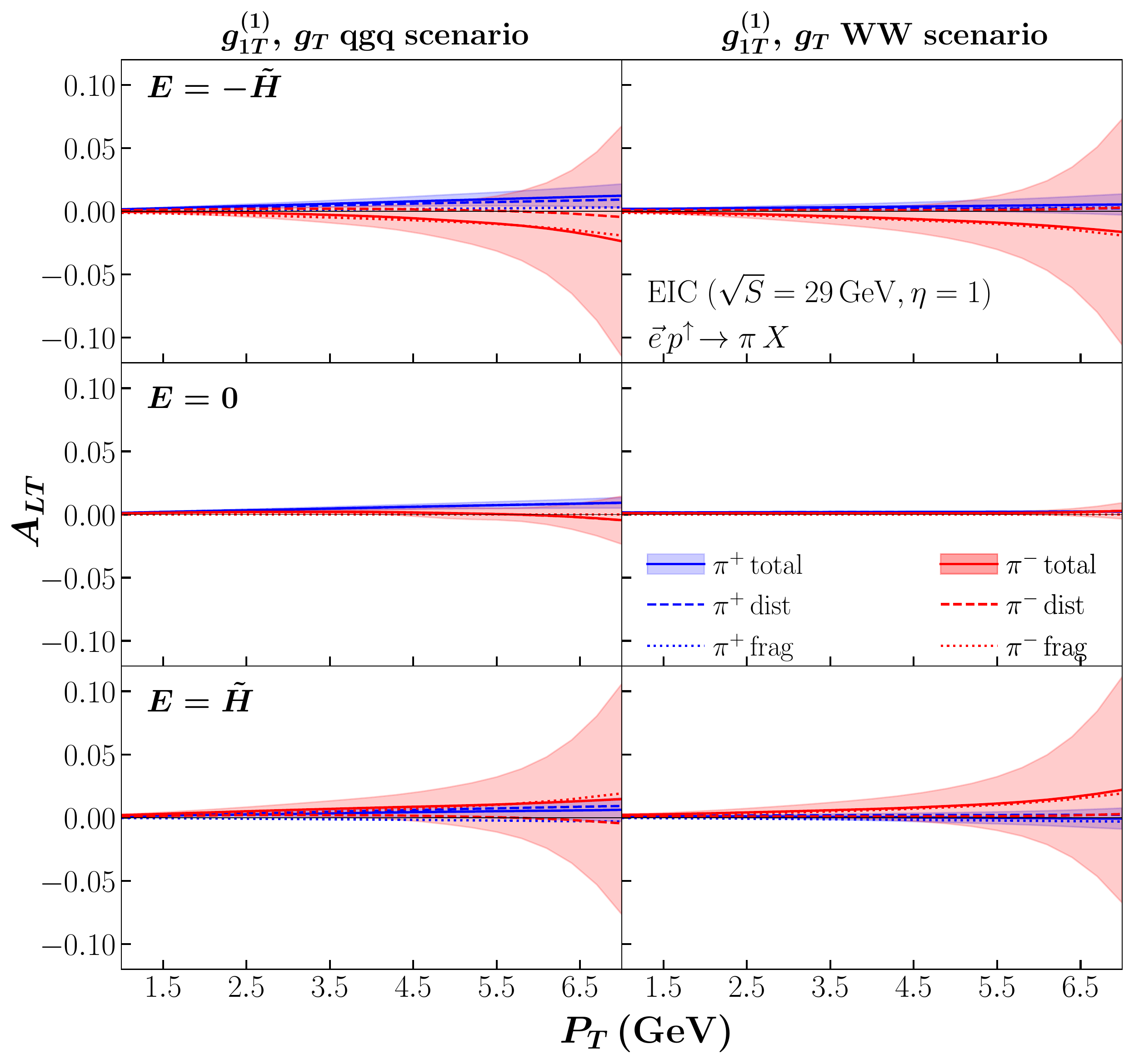}\vspace{-0.2cm}
\caption{Predictions for $A_{LT}$ vs.~$P_T$ in $\vec{e}\,p^\uparrow\!\to \pi\,X$ for low-energy EIC kinematics at slightly forward rapidity ($\sqrt{S}=29\,{\rm GeV},\eta=1$).   The description is the same as the Fig.~\ref{f:JLab6} caption.\vspace{-0.3cm}} 
\label{f:EIC29_eta1}
\end{figure}

\begin{figure}[h!]
\includegraphics[width=0.625\textwidth]{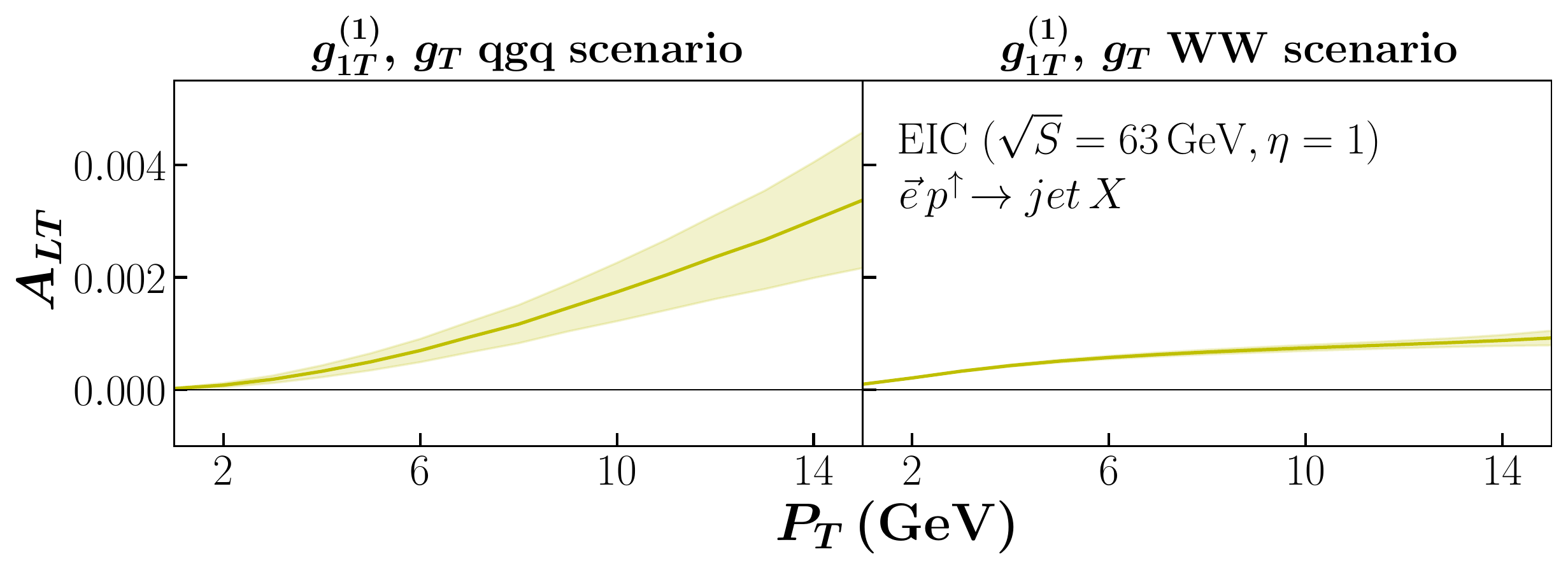}\vspace{-0.2cm}
\caption{Predictions for $A_{LT}$ vs.~$P_T$ in $\vec{e}\,p^\uparrow\!\to jet\,X$ for higher-energy EIC kinematics at slightly forward rapidity ($\sqrt{S}=63\,{\rm GeV},\eta=1$).   \vspace{-0.3cm}} 
\label{f:EIC63jet_eta1}
\end{figure}

In this appendix we include the plots discussed in Secs.~\ref{s:JLab6}, \ref{s:JL_COM_EIC} for JLab6 (Fig.~\ref{f:JLab6}), JLab12 (Fig.~\ref{f:JLab12}), COMPASS (Fig.~\ref{f:COMPASS}), low-energy EIC for pion production at midrapidity (Fig.~\ref{f:EIC29_eta0}) and slightly forward rapidity (Fig.~\ref{f:EIC29_eta1}), and higher-energy EIC for jet production at slightly forward rapidity (Fig.~\ref{f:EIC63jet_eta1}).


\clearpage

\section{Proton-Proton Results}
\label{s:app_b}

\begin{figure}[h!]
\includegraphics[width=0.625\textwidth]{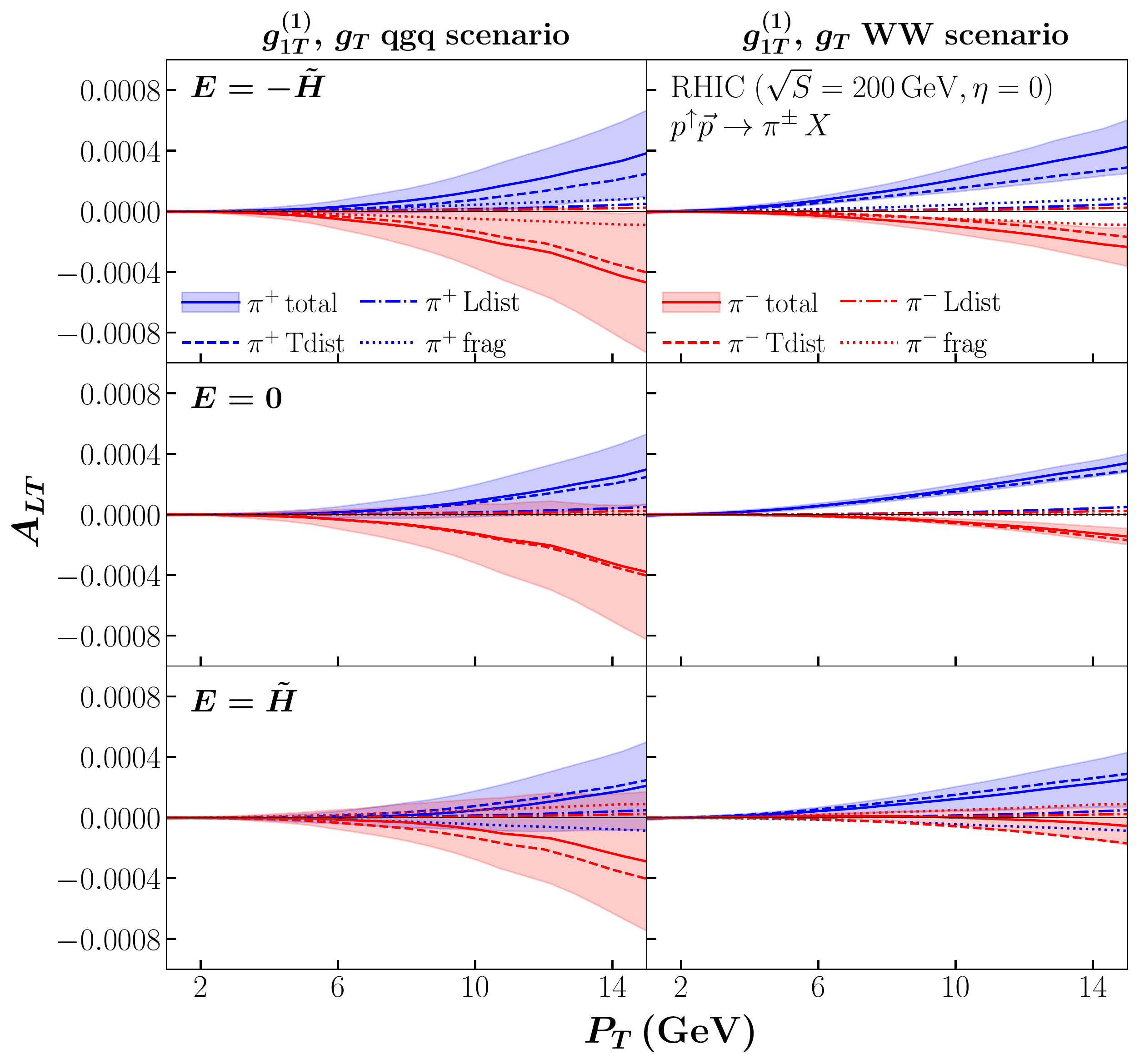}\vspace{-0.2cm}
\caption{Predictions for $A_{LT}$ vs.~$P_T$ in $p^\uparrow \vec{p}\! \to \pi^\pm\,X$ for RHIC kinematics at midrapidity ($\sqrt{S}=200\,{\rm GeV},\eta=0$).  The left column is for the qgq scenario for $g_{1T}^{(1)}(x)$, $g_T(x)$ and the right is for the WW scenario (see Sec.~\ref{s:numerics} for more details).  The first row is for the case $E(z)=-\tilde{H}(z)$, the second for $E(z)=0$, and third for $E(z)=\tilde{H}(z)$.  The solid curve gives the average total asymmetry (with $68\%$ C.L.~error band), while the dashed (dashed-dotted, dotted) curves give the average individual contribution from the transverse distribution (longitudinal distribution, fragmentation) term. \vspace{-0.3cm}} 
\label{f:RHIC_mid_pipm}
\end{figure}

\begin{figure}[h!]
\includegraphics[width=0.625\textwidth]{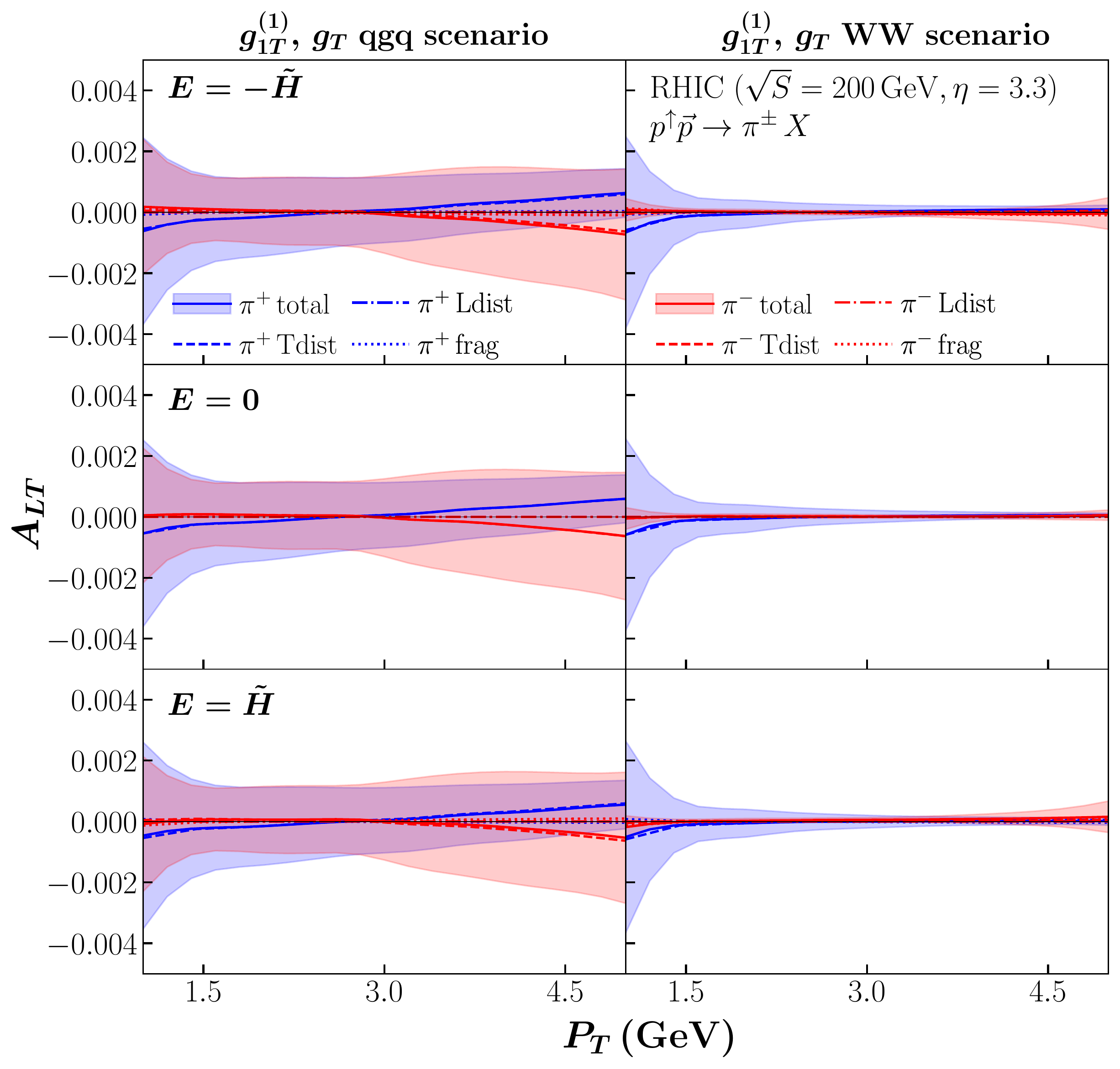}\vspace{-0.2cm}
\caption{Predictions for $A_{LT}$ vs.~$P_T$ in $p^\uparrow \vec{p}\! \to \pi^\pm\,X$ for RHIC kinematics at forward rapidity ($\sqrt{S}=200\,{\rm GeV},\eta=3.3$). The description is the same as the Fig.~\ref{f:RHIC_mid_pipm} caption.\vspace{-0.3cm}} 
\label{f:RHIC_forward_pipm}
\end{figure}

\begin{figure}[h!]
\includegraphics[width=0.625\textwidth]{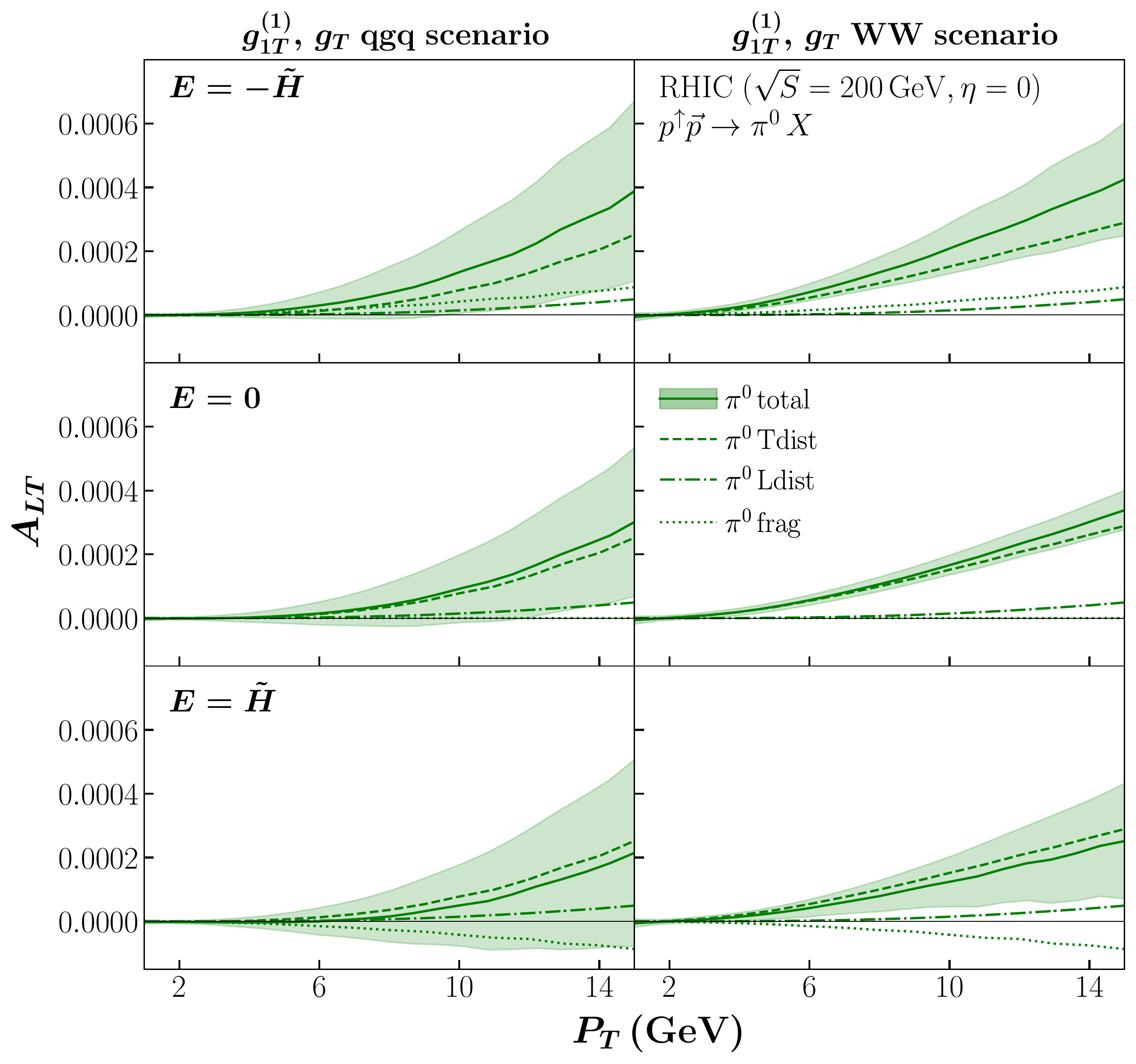}\vspace{-0.2cm}
\caption{Predictions for $A_{LT}$ vs.~$P_T$ in $p^\uparrow \vec{p}\! \to \pi^0\,X$ for RHIC kinematics at midrapidity ($\sqrt{S}=200\,{\rm GeV},\eta=0$). The description is the same as the Fig.~\ref{f:RHIC_mid_pipm} caption.\vspace{-0.3cm}} 
\label{f:RHIC_mid_pi0}
\end{figure}

\begin{figure}[h!]
\includegraphics[width=0.625\textwidth]{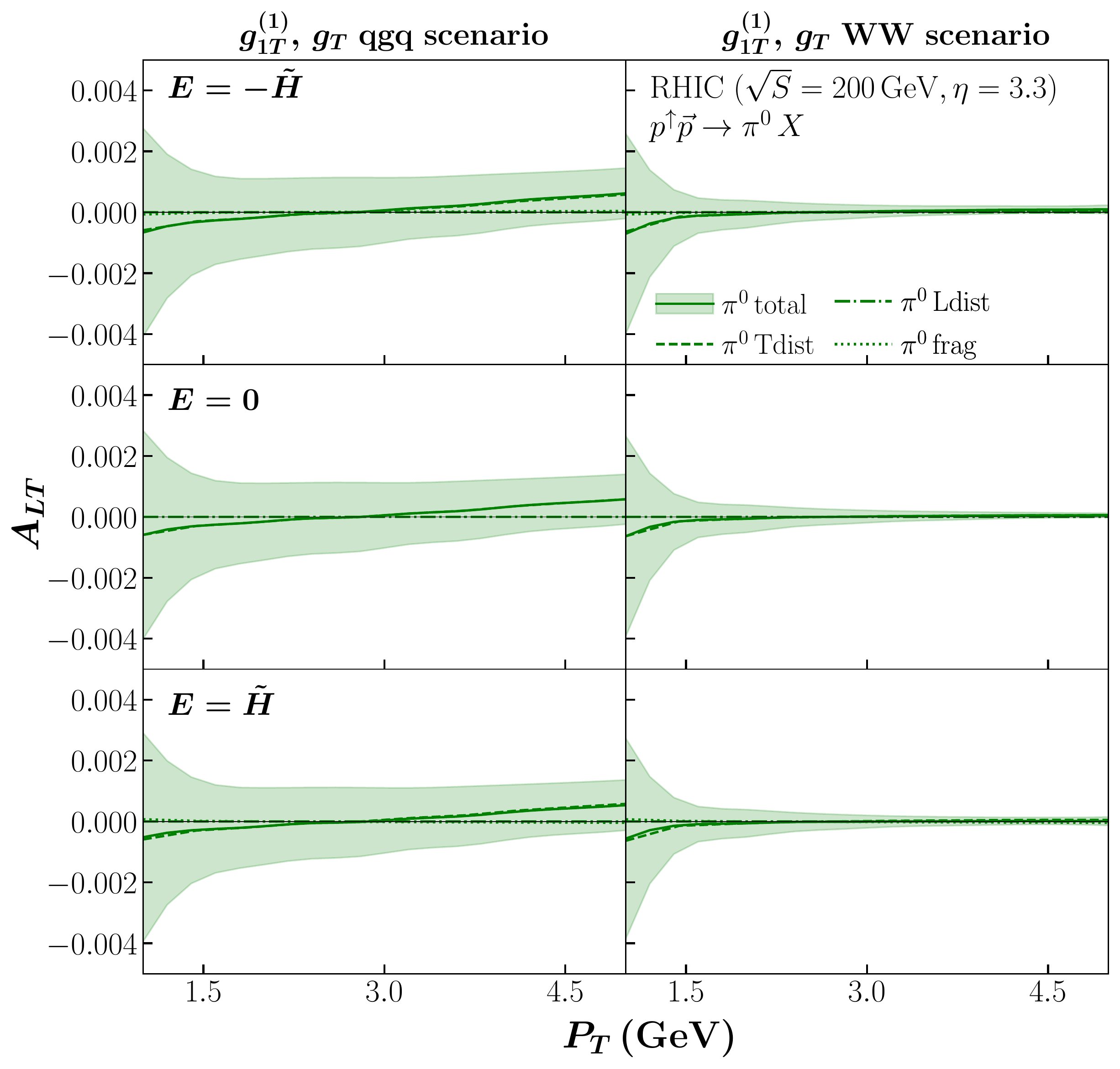}\vspace{-0.2cm}
\caption{Predictions for $A_{LT}$ vs.~$P_T$ in $p^\uparrow \vec{p}\! \to \pi^0\,X$ for RHIC kinematics at forward rapidity ($\sqrt{S}=200\,{\rm GeV},\eta=3.3$). The description is the same as the Fig.~\ref{f:RHIC_mid_pipm} caption.\vspace{-0.3cm}} 
\label{f:RHIC_forward_pi0}
\end{figure}

\begin{figure}[h!]
\includegraphics[width=0.63\textwidth]
{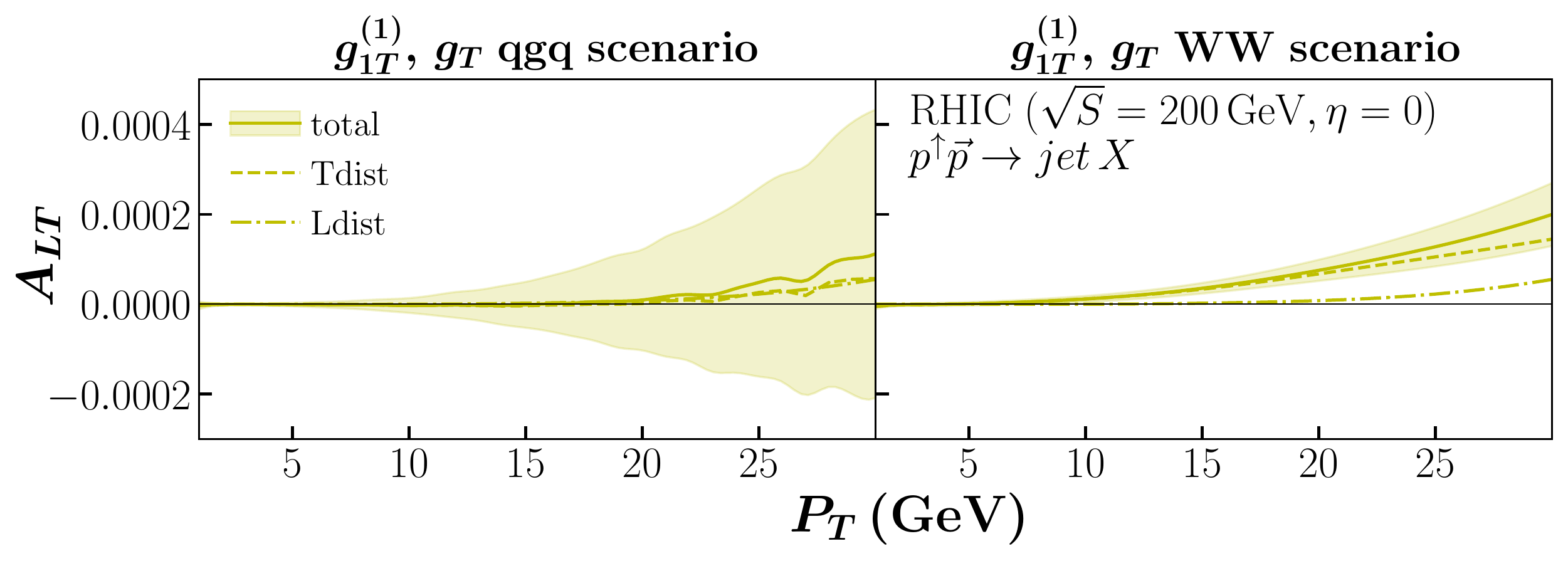}\\
\hspace{0.125cm}\includegraphics[width=0.632\textwidth]{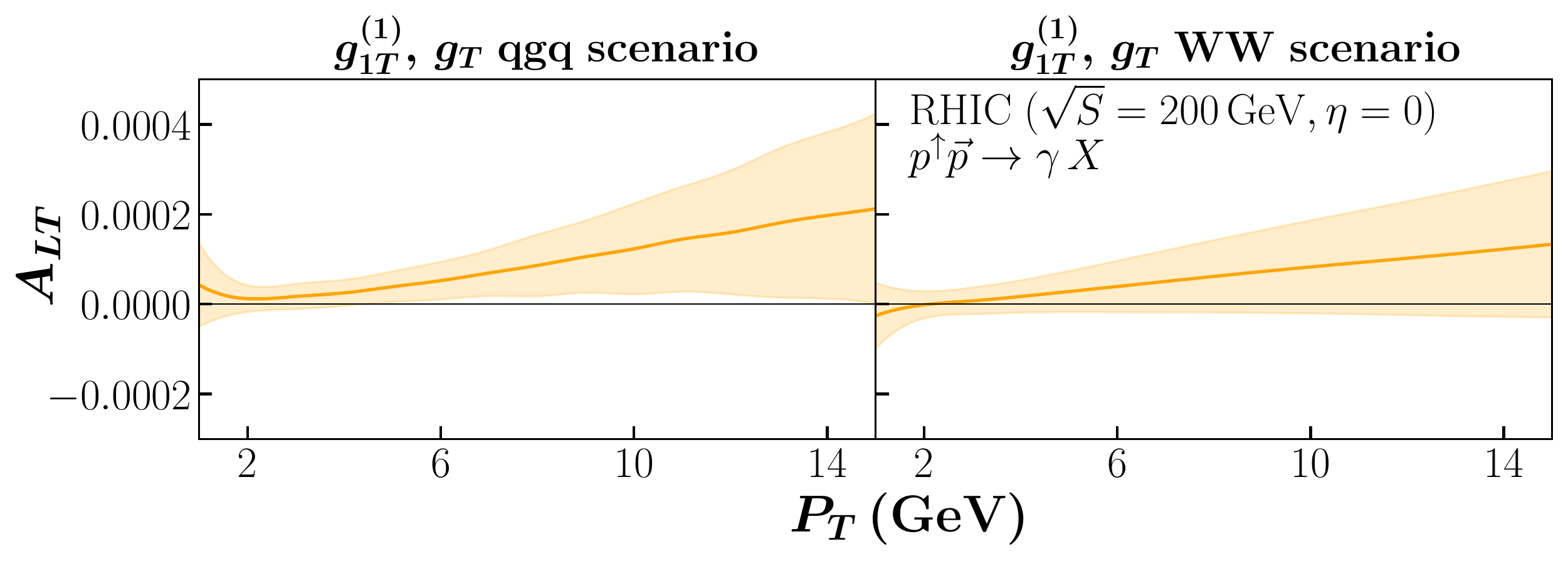}
\vspace{-0.2cm}
\caption{Predictions for $A_{LT}$ vs.~$P_T$ in $p^\uparrow \vec{p}\! \to jet\,X$ (top) and $p^\uparrow \vec{p}\! \to \gamma\,X$ (bottom) for RHIC kinematics at midrapidity ($\sqrt{S}=200\,{\rm GeV},\eta=0$). The left column is for the qgq scenario for $g_{1T}^{(1)}(x)$, $g_T(x)$ and the right is for the WW scenario (see Sec.~\ref{s:numerics} for more details).  (There is no fragmentation term, so one does not have to consider different scenarios for $E(z)$.)  The solid curve gives the average total asymmetry (with $68\%$ C.L.~error band), while for the jet case the dashed (dashed-dotted) curves give the average individual contribution from the transverse distribution (longitudinal distribution) term. (There is no analytical result for the longitudinal distribution term for the photon case, so the total and transverse distribution piece are one in the same.) \vspace{-0.3cm}} 
\label{f:RHIC_mid_jet_gam}
\end{figure}

In this appendix we include the plots discussed in Sec.~\ref{s:RHIC} for RHIC for $\pi^\pm$ at midrapidity (Fig.~\ref{f:RHIC_mid_pipm}) and forward rapidity (Fig.~\ref{f:RHIC_forward_pipm}), for $\pi^0$ production at midrapdity (Fig.~\ref{f:RHIC_mid_pi0}) and forward rapidity (Fig.~\ref{f:RHIC_forward_pi0}) , and for jet or photon production at midrapidity (Fig.~\ref{f:RHIC_mid_jet_gam}).

\end{document}